\documentclass[12pt]{article}
\usepackage[utf8x]{inputenc}
\usepackage{hyperref}
\usepackage{scicite}
\usepackage{enumerate}
\usepackage{float}
\usepackage{times}
\usepackage{amssymb,latexsym,amsmath,graphicx,color}
\usepackage{multirow}

\newenvironment{sciabstract}{%
\begin{quote} \bf}
{\end{quote}}

\newcounter{lastnote}

\title{\bf Ultrastrong plasmon-phonon coupling via epsilon-near-zero nanocavities}

\author
{Daehan Yoo,$^{1 \;\dagger}$
Fernando de Le\'on-P\'erez,$^{2,3 \;\dagger}$  
In-Ho Lee,$^{1}$
Daniel A. Mohr,$^{1}$\\
Matthew Pelton,$^{4}$
Markus B. Raschke,$^{5}$
Joshua D. Caldwell,$^{6}$\\
Luis Mart\'{i}n-Moreno,$^{3\ast}$ 
and Sang-Hyun Oh$^{1\ast}$\\
\\
\normalsize{$^{1}$Department of Electrical and Computer Engineering, University of Minnesota,}\\
\normalsize{200 Union St. S.E., Minneapolis, MN, USA,}\\
\normalsize{$^{2}$Centro Universitario de la Defensa de Zaragoza,}\\
\normalsize{Ctra. de Huesca s/n, E-50090 Zaragoza, Spain,}\\
\normalsize{$^{3}$Instituto de Ciencia de Materiales de Aragón and}\\ 
\normalsize{Departamento de Física de la Materia Condensada,}\\
\normalsize{CSIC-Universidad de Zaragoza, E-50009 Zaragoza, Spain,}\\
\normalsize{$^{4}$Department of Physics, University of Maryland, Baltimore County, Baltimore, MD, USA,}\\
\normalsize{$^{5}$Department of Physics and JILA, Department of Chemistry,}\\ 
\normalsize{University of Colorado, Boulder, CO, USA,}\\
\normalsize{$^{6}$Department of Mechanical Engineering, Vanderbilt University, Nashville, TN, USA.}\\
\\
\normalsize{{$^\ast$E-mail: \textcolor{blue}{lmm@unizar.es} and \textcolor{blue}{sang@umn.edu}}}\\
\normalsize{$^\dagger$These authors contributed equally to this work}} 

\date{}

\begin{document} 


\baselineskip24pt


\maketitle 


\begin{sciabstract}
Vibrational ultrastrong coupling (USC), where the light-matter coupling strength is comparable to the vibrational frequency of molecules, presents new opportunities to probe the interactions of molecules with zero-point fluctuations, harness cavity-enhanced chemical reactions, and develop novel devices in the mid-infrared regime. Here we use epsilon-near-zero nanocavities filled with a model polar medium (SiO$_2$) to demonstrate USC between phonons and gap plasmons. We present classical and quantum mechanical models to quantitatively describe the observed plasmon-phonon USC phenomena and demonstrate a splitting of up to 50\% of the resonant frequency. Our wafer-scale nanocavity platform will enable a broad range of vibrational transitions to be harnessed for USC applications.

\end{sciabstract}



Traditionally, two regimes of light-matter interactions have been considered: weak coupling (WC), when losses exceed the light-matter coupling strength, and strong coupling (SC), when the coupling strength dominates \cite{TormaRPP15,EbbesenACR16}.
For weak-coupling phenomena such as Purcell effect \cite{PeltonNPh19} and Fano interference \cite{LukyanchukNM10}, the coupled systems exchange energy on a time scale slower than the decay rates. In contrast, within the SC regime, the oscillators exchange their energy reversibly and coherently over an extended time frame that is longer than the decay rates\cite{YoshieN04,ReithmaierN04,AokiN06,liu2015strong,BenzS16,SanthoshNC16,runnerstrom2018polaritonic,LengNC18,ParkSA19} -- a prerequisite for quantum information processing\cite{SchoelkopfN08}. Furthermore, strongly-coupled hybridized modes exhibit distinct energy levels and characteristics modified from those of the bare constituents, leading to novel phenomena such as the modification of chemical reaction rates \cite{EbbesenACR16, dunkelberger2016modified, MunkhbatN18} and ground-state reactivity \cite{ThomasS19}.

When the normalized coupling strength, $\eta$ (defined as the ratio of the vacuum Rabi splitting, $\Gamma$, to the resonance frequency), of the system exceeds $0.1 \sim 0.2$, even more exotic phenomena can occur. In this ultrastrong coupling (USC) regime \cite{TormaRPP15,EbbesenACR16,KockumNRP19,FornRMP19}, some of the standard approximations that are valid for WC and SC, such as the rotating-wave approximation, are expected to break down. Furthermore, transitioning from SC to USC implies that the hybrid mode exhibits substantially more oscillations between light and matter states prior to decay, and such fast and efficient interactions in USC can enable novel ultrafast devices\cite{KockumNRP19,RomeroPRL12}. Another striking phenomenon predicted in the USC regime, resulting from the antiresonant term in the light-matter coupling equation, is the possibility to extract virtual photons from the modified ground state via a dynamic Casimir effect\cite{CiutiPRB05,KockumNRP19}  Also, USC between light and molecules can modify or enhance chemical reactions beyond what is possible in the SC regime \cite{KockumNRP19}. 

USC has been demonstrated using photochromic molecules\cite{schwartz2011reversible}, circuit QED systems\cite{niemczyk2010circuit}, molecular liquids\cite{george2016multiple}, and 2D electron gases \cite{scalari2012ultrastrong}. However, for technologically important mid-infrared (MIR) applications such as surface-enhanced vibrational spectroscopy\cite{NeubrechPRL08,MullerACSPh18}, thermal emission and signature control, and modified heat transfer, it is desirable to attain vibrational USC using solid-state materials. While vibrational SC in the MIR has been demonstrated \cite{AutoreLSA18,LatherACIE19,ThomasS19}, MIR USC has been difficult to achieve, because the oscillator strengths for vibrational modes originating from lattice ionic motions are much weaker than those from excitons or artificial atoms in circuit QED. 

Here we use tightly coupled gap plasmons and phonons in coaxial nanocavities whose ultranarrow gaps are filled with a phononic material to attain vibrational USC in the MIR. Close to the cutoff frequency of the TE$_{11}$-like mode, the coaxial nanocavities exhibit strong transmission resonances and field enhancements, which result from the zeroth-order Fabry-Perot mode\cite{BaidaPRB06} or alternatively, the effective epsilon-near-zero (ENZ) effect\cite{AluPRB08, YooNL16,liberal2017near}. We couple this nano-coax ENZ mode to the lattice vibrations of SiO$_2$ and demonstrate USC exceeding 50\% of the vibrational transition frequency.

\begin{figure}[H]
\centering
\includegraphics[scale=0.5]{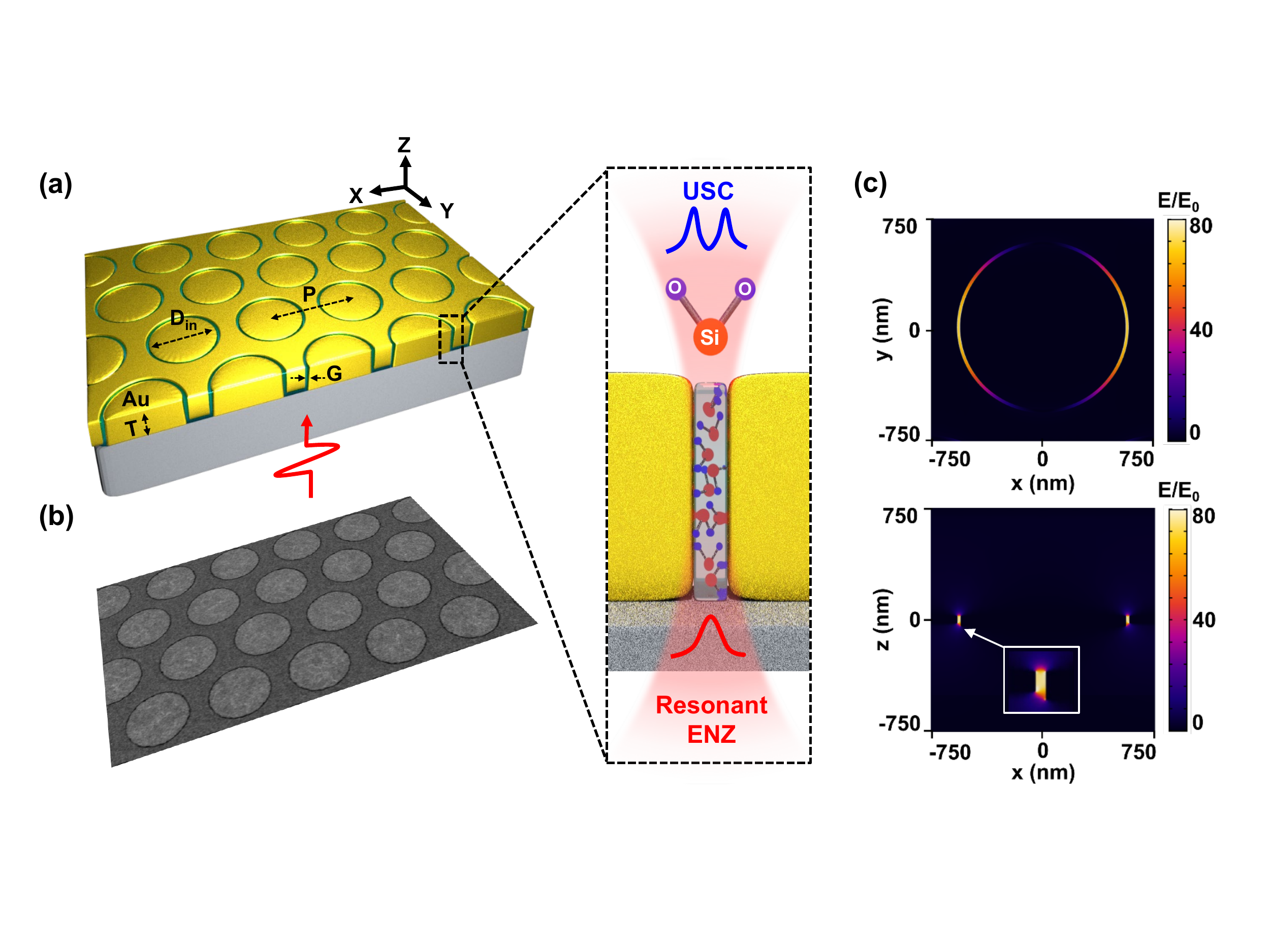}
\caption{Wafer-scale resonant ENZ nanocavity platform for USC. (a) Geometrical parameters and illustration of SiO$_2$ phonons interacting with the epsilon-near-zero (ENZ) mode of the nanocavity. (b) Scanning electron micrograph (SEM) of coaxial nanocavities with 21 nm gap, 790 nm diameter, 1190 nm period, and 80 nm Au thickness. (c) Computed electric field profiles of a coaxial nanocavity showing a lateral section (upper) and a vertical section (lower) of the ENZ transmission resonance.}
\label{fig:Fig1}
\end{figure} 

In our USC platform, SiO$_2$-filled coaxial nanoapertures defined into a metal (Au) film are hexagonally arranged with varied gap size ($G$), diameter ($D$), and lattice periodicity ($P$) (Figs. 1a,b). Unlike conventioal approaches of selectively etching annular gaps in metal films, we adopt a novel fabrication approach called atomic layer lithography\cite{YooNL16} to create dielectric-filled nanogaps in metal films. After standard photolithography to define gold disk arrays on a silicon wafer, SiO$_2$ films are conformally grown via atomic layer deposition (ALD) on the exposed surfaces and sidewalls followed by the deposition of the metal cladding layer and planarization via glancing-angle ion milling. This batch process can produce wafer-sized arrays of coaxial apertures with gap size down to ~1 nm, limited by the critical ALD growth step. 

For very narrow gaps, TE$_{11}$ coax modes have a strong plasmonic character. Such modes are characterized by an effective dielectric constant, which is approximately zero at cutoff.  This resonance can be shifted toward longer wavelengths without sacrificing confinement by increasing the coax diameter without changing the gap width. In addition, the associated spatially uniform optical field (Fig. 1c lower panel), due to the very long wavelength at near-zero-permittivity, provides efficient coupling to the material and its resonances within the gaps. We also note that the ENZ resonance is a single-aperture effect, so the existence of an array is not required for the transmission resonance we utilize (Sec. S1). 

Our coaxial nanocavities were designed to sweep the bare ENZ resonance across the Reststrahlen band defined by the transverse optical (TO) and longitudinal optical (LO) phonons of SiO$_2$ (Fig. 2a). Transmission spectra through the nanocavity arrays were measured by far-field Fourier transform infrared (FTIR) spectroscopy over a large-area (5 mm $\times$ 5 mm) chip containing arrays of coaxial nanocavities (diameter from 430 nm to 1120 nm in 30 nm steps). The normalized transmission spectra for coaxial nanocavities filled with 21 nm-thick SiO$_2$ are plotted in Fig. 2b. The observed transmission peak is clearly split, with an anticrossing behavior characteristic of SC. Coaxial nanocavities at the ENZ condition show extraordinary optical transmission\cite{garcia2010light}, since incident light with wavelengths 5-13 $\mu$m can pass through the annular gaps that are over thousand times narrower than the wavelength, with the measured transmittance (normalized by the total chip area) as high as 60\%.


\begin{figure}[H]
\centering
\includegraphics[scale=0.6]{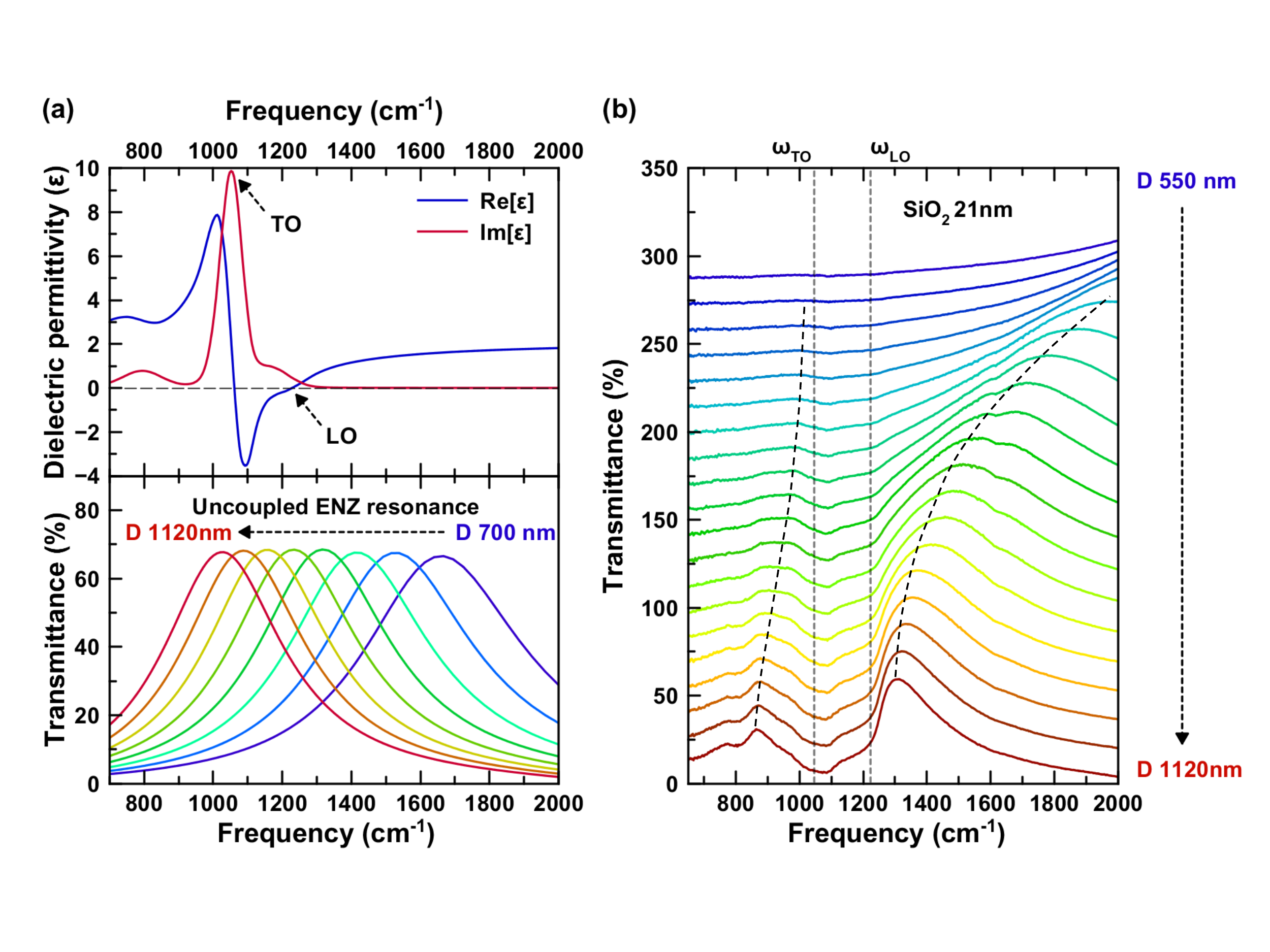}
\caption{Rabi splittings due to ultrastrong phonon-plasmon coupling. (a) Dielectric function \cite{KischkatAO12} of SiO$_2$ (top) and ENZ transmission resonances (bottom) of the coaxial nanoapertures ($G=21$ nm, diameter range: 700 to 1120 nm), simulated with a frequency-independent dielectric constant $\epsilon_\infty$ inside the hole. This calculation shows that the bare cavity resonance sweeps the Reststrahlen band for the considered diameters (Sec. S2). (b) Normalized experimental transmission spectra of SiO$_2$-filled coaxial Au nanocavities, for diameters ranging from $D =$ 550 nm to $D =$ 1120 nm.  The coaxial gap is 21 nm. Black dashed lines are guides for the eye indicating the lower and upper polariton branches.}
\label{fig:Fig2}
\end{figure}

Although these results suggest that the SiO$_2$ phonons and cavity photons are strongly coupling, the split transmission peaks could alternatively be interpreted as the result of a single broad transmission resonance being quenched in a central frequency range by the strong vibrational absorption close to $\omega_{TO}$. To clarify the origin of the split transmission peaks, we perform a theoretical analysis of the coupled system.

Electromagnetic propagation in waveguides is usually based on a description where the phononic degrees of freedom have been integrated out, providing an effective medium $\epsilon(\omega)$ for the propagation of the photon (Sec. S3). However, in order to determine whether the system is in the SC regime, it is essential to retain both phononic and photonic degrees of freedom. We therefore consider a given waveguide mode M, characterized by a wavevector $k$ along the waveguide axes (in this case, M=TE$_{11}$, the fundamental mode of the coaxial waveguide). When the aperture is filled with a uniform dielectric constant $\epsilon_\infty$ (originating from coupling to electronic degrees of freedom), the electric field, $\mathbf{E}=E\mathbf{E}_M$, satisfies the wave equation
\begin{eqnarray}
\label{eq:webare}
\nabla \times \nabla \times E \mathbf{E}_M   -\epsilon_\infty \frac{\omega^2_k}{c^2} E \mathbf{E}_M =0, 
\end{eqnarray}
where $E$ is the field amplitude and $\mathbf{E}_M$ is a normalized transverse solution of Maxwell equations (Sec. S4).
When the phononic material fills the waveguide, we assume a local relation (valid for non-dispersive vibrational modes) between $\mathbf{E}$ and the relative displacement of ions, $\mathbf{x}$, following \cite{bornhuang}. Then, for a given mode profile, $\mathbf{E}(\mathbf{r})=E \; \mathbf{E}_M (\mathbf{r},k)$ and $\mathbf{x}(\mathbf{r})=x \; \mathbf{E}_M (\mathbf{r},k)$, with 
\begin{eqnarray}
\label{eq:snl}
 \ddot{x} &=& \gamma_{11} x+ \gamma_{12}E, \\
\label{eq:polf}
P &=& \gamma_{12} x+ \gamma_{22} E, 
\end{eqnarray}
where $P$ is the amplitude of polarization vector $\mathbf{P}=P \; \mathbf{E}_M$, $\gamma^2_{12}=\omega^2_p\epsilon_\infty/4\pi$ is the coupling constant, $\omega^2_p=\omega^2_{LO}-\omega^2_{TO}$, $\gamma_{11}=-\omega^2_{TO}$, and  $\gamma_{22}=(\epsilon_\infty-1)/4\pi$ (see Sec. S3). Note that $\gamma_{22}$ incorporates the effect of the electronic resonances excited at higher frequencies. The wave equation becomes, in this case,
\begin{eqnarray}
\label{eq:wefilled}
\nabla \times \nabla \times E \mathbf{E}_M  - \frac{\omega^2}{c^2} \left( E+4\pi P\right) \mathbf{E}_M=0. 
\end{eqnarray}

By inspection, we find that $\mathbf{E}_M(\mathbf{r},k)$ satisfies Eq. (\ref{eq:wefilled}) if $\omega^2(E+4\pi P)=\omega_k^2 E$ (strictly speaking, we are neglecting the variation between $\omega$ and $\omega_k$ in the impedance of the metal surrounding the waveguide; for a deeper analysis of how this affects the cutoff condition, see Secs. S1 and S5). This condition, together with Eqs. (\ref{eq:snl}) and (\ref{eq:polf}) can be expressed in a matrix form, stating that $\mathbf{E}_M(\mathbf{r},k)$ is still a solution of Maxwell equations but at a frequency $\omega$ satisfying 
\begin{eqnarray}
\label{eq:2x2}
\left( \begin{array}{cc}
\omega^2-\omega^2_{TO} &  \omega \; \omega_p  \\
\omega \; \omega_p &  \omega^2 -\omega^2_k
\end{array} \right) \cdot \left( \begin{array}{c}
\omega \; x \\ \sqrt{\epsilon_\infty/4\pi}  E
\end{array} \right) &=& 0 .  
\end{eqnarray}
(See Sec. S6 for a detailed derivation and a discussion of the effects of losses)
Note that the simpler coupled harmonic oscillator model that is customarily used for fitting experimental data (see Sec. S7) does not show the $\omega$-dependence of the off-diagonal terms and therefore cannot describe the polaritonic branches correctly (cf. Figs. 3a and S1).  

We have used a classical approach here, and it may be questioned whether this approach provides valid results in the USC regime, where quantum-mechanical effects are unavoidable, and off-resonant terms may be important ({\em i.e.}, the rotating-wave approximation cannot necessarily be applied \cite{HopfieldPR58,QuattropaniNC86,CiutiPRB05,KenaCohenAOM13}.
We therefore applied the canonical procedure of second quantization to the classical Hamiltonian (see Sec. S9). For the sake of convenience, absorption was neglected and the geometry of the system was simplified to a coaxial resonator with perfect-electric-conductor walls.  
We obtain a Hopfield-like Hamiltonian \cite{HopfieldPR58} for interacting photons and phonons: 
$H=H_{photon}+H_{phonon}+H_{int}$, with
\begin{eqnarray}
\label{eq:Hopfield}
H_{photon} &=& \sum_m \hbar \omega_m \left( a^+_m a_m+\frac{1}{2}\right),  \\
H_{phonon} &=& \sum_m \hbar \omega_{TO} \left( b^+_m b_m+\frac{1}{2}\right),  \\
H_{int} &=& \sum_m \hbar \left[i C_m \left(a^+_m b_m-a_m  b^+_m \right)+D_m  \left(2a^+_k a_m+1 \right) \right.  \\
& & \left. +iC_m\left(a_m b_m -a^+_m b^+_m \right) +D_m  \left(a_m a_m+a^+_m a^+_m \right)  \right]. 
\end{eqnarray}
where $a^+$ ($a$) and $b^+$ ($b$) are the creation (annihilation) operators for photons and phonons, respectively, the sum is over modes $m$, and
\begin{eqnarray}
\label{eq:CD}
C_m &=& \frac{\omega_p}{2} \sqrt{\frac{\omega_{TO}}{\omega_m}}, \;\;
D_m = \frac{\omega^2_p}{4 \omega_m}. 
\end{eqnarray}
Diagonalizing the full Hamiltonian gives eigenfrequencies identical to those obtained from Eq. \ref{eq:2x2} (if absorption is neglected). In other words, the classical approach can be used to compute the polaritonic branches exactly when the system enters into the USC regime.  The same behavior is valid for an arbitrary number of vibrational modes in the polar material (see Secs. S8 and S10).

Eq. \ref{eq:2x2} is thus one of the central results of our work. It shows that the dynamics between the vibrations and the plasmonic field in nanocavities are controlled by a $2\times 2$ matrix similar to that used to discuss SC. The main difference is that the off-diagonal elements acquire a linear dependence on frequency. This does not have a large effect when analyzing anti-crossings, which focus on a narrow frequency range, but it is essential to understand the presence of an energy gap in the spectrum i.e., the fact that the low-frequency asymptote of the upper polariton and the high-frequency asymptote of the lower polariton do not coincide (see Fig. 3a). 


\begin{figure}[H]
\centering
\includegraphics[scale=0.5]{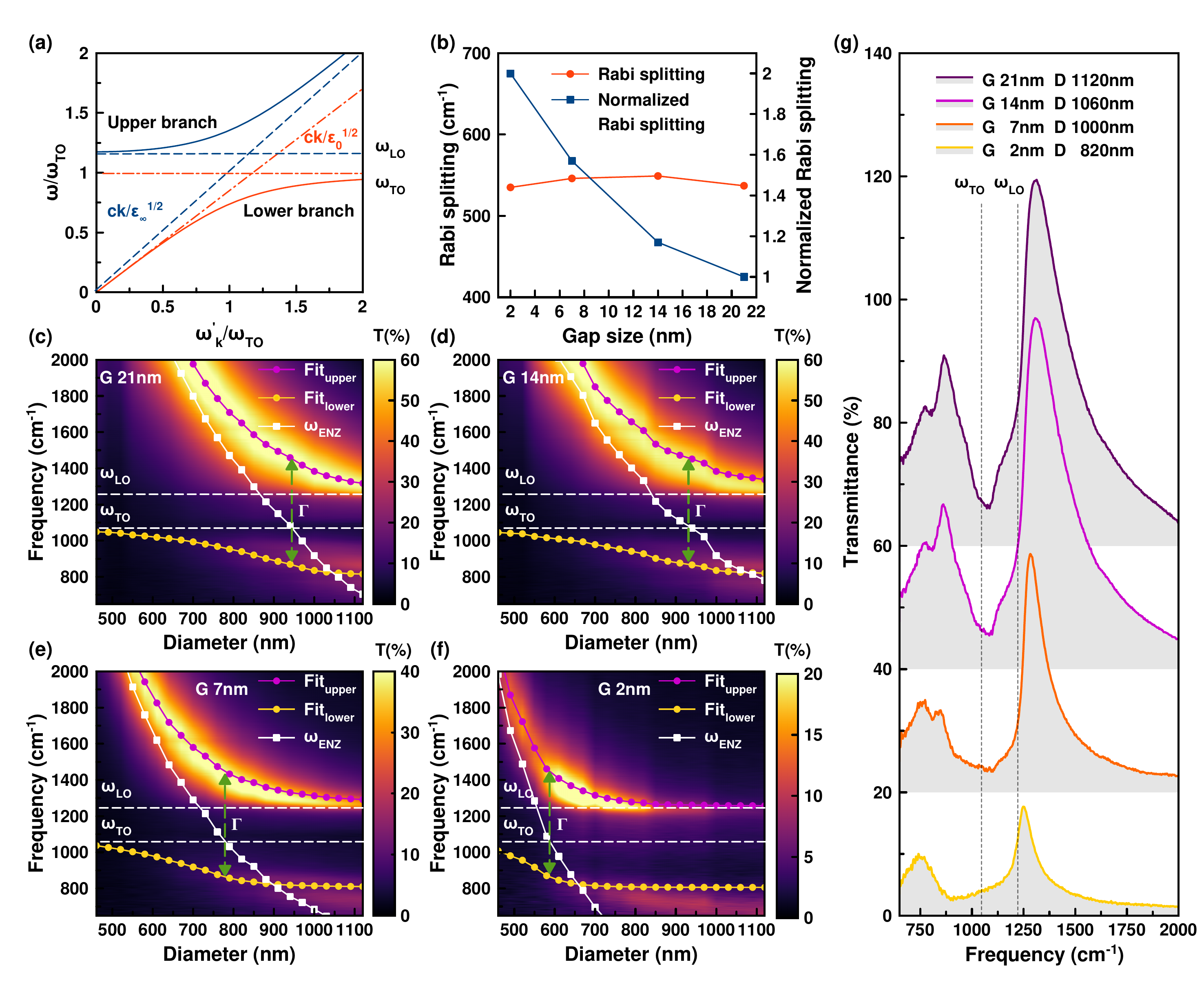}
\caption{Dispersion mapping and validation of theoretical model. (a) Dispersion relation of bulk phonon polaritons in SiO$_2$ calculated by Eq. \ref{eq:2x2} assuming that a single vibrational mode is excited. (b) Measured Rabi splitting and normalized Rabi splitting vs. gap size. (c-f) Dispersion maps of experimentally measured transmission spectra from coaxial apertures filled with SiO$_2$ film thicknesses of (c) 21 nm, (d) 14 nm, (e) 7 nm, and (f) 2 nm. Analytical fitting of polaritonic resonances using Eq. (S14) (purple (upper) and yellow (lower) circle dots with a solid line). $\omega_{ENZ}$ (white square dots with a solid line) is used as a fitting parameter and indicates the resonance of the ENZ mode uncoupled with polar phonons of SiO$_2$. Rabi splitting ($\Gamma$) is measured at the intersection between the uncoupled ENZ resonance and $\omega_{TO}$ (when the detuning is zero). (g) Transmission spectra measured when the bare ENZ mode crosses the Reststrahlen band, showing distinct Rabi splitting, sharp peaks, and extraordinary optical transmission (peak transmittance of 60\% for 21 nm gap coax). Curves are offset along the vertical axis for clarity.}
\label{fig:Fig3}
\end{figure}

We can therefore analyze the dependence of the ENZ resonant frequencies on the coax diameter using Eq. \ref{eq:2x2} (Fig. 3 c-g). As $\omega_p$ is independent of the hole shape and size, in this SC condition the ENZ resonances only depend on the bare ENZ frequency, $\omega_{ENZ}(D)$, obtained from the upper polariton band (see Sec. S12). The model provides the estimate for the lower polariton band. This procedure was followed because the lower polariton is more affected by additional vibrational resonances that have not been considered in this simple analysis. 


For each gap size, we estimate the Rabi splitting of the ENZ mode as the frequency difference between upper and lower polariton branches when $\omega_{ENZ}(D)=\omega_{TO}$; results are summarized in Table 1. Although the linewidth of the uncoupled ENZ resonance at zero-detuning cannot be measured, we take advantage of its small dependence on hole diameter, and estimate its value from ENZ resonances occurring at higher frequencies (where they do not couple with phonon modes). As shown in Table 1, all cavities exhibit mode splitting larger than the average of the linewidths of the bare cavity resonance and the TO phonon ($\Gamma/\gamma_{avg} > 1$), indicating that the system is in the SC regime.


Moreover, the coupling strength $\eta = \Gamma / \omega_{TO} > 0.5$, regardless of gap width, indicating that the systems are also in the USC regime \cite{KockumNRP19}. The independence of $\eta$ on the gap width, and thus of the mode volume $V$, can be understood by considering that the coupling constant per emitter is proportional to $1/\sqrt{V}$, but the number of dipoles coherently coupled to the mode, $N \propto V$. Since $\Gamma \propto \sqrt{N}$, it is independent of $V$, and depends only on the density of molecules, $N/V$ (see Sec. S3).

Interestingly, linewidth narrowing of the coupled modes is observed in the SiO$_2$-filled coaxes, compared to the bare cavity resonances (Figs. S2-S3). The hybridized modes exhibit plasmon- and phonon-like properties simultaneously, resulting in linewidths that are the average of the plasmon and phonon modes. While uncoupled ENZ modes usually exhibit a $Q$-factor of $2\sim3$, the hybrid modes generated in the vicinity of the asymptotic limits of the LO phonon exhibits $Q > 5$ for 2 nm-gap coaxial cavities (Table S1), where the coupling strength is maximum.  

\begin{table}[H]
\begin{center}
\begin{tabular}{c|c|c|c|c} \hline \hline
SiO$_2$ gap (nm) & 2 & 7 & 14 & 21 \\ \hline
Bare ENZ mode linewidth (cm$^{-1}$)	& 453	& 619	& 844	& 990 \\ 
Average of bare ENZ and TO phonon linewidths, $\gamma_{avg}$ (cm$^{-1}$)	& 265 & 348	& 481 & 534 \\
Rabi splitting, $\Gamma$ (cm$^{-1}$) &	534	& 546	& 549	& 537 \\
Normalized Rabi splitting, $\Gamma/\gamma_{avg}$: & {2.02} &	{1.57} &	{1.14} &	{1.01} \\
Coupling strength $\eta = \Gamma/\omega_{TO}$ & 0.508 & 0.519	& 0.522 & 0.510 \\
Blue shift of the central frequency$/\omega_{TO}$ (\%) & 6 & 8 & 
10 & 11 \\ \hline \hline
\end{tabular}
\end{center}
\caption{Rabi splitting, coupling strength, and blue shift of the central frequency.}
\label{table:Rabi}
\end{table} 

In summary, we have constructed wafer-scale resonant coaxial nanocavities filled with SiO$_2$ and achieved a coupling strength ($\eta=\Gamma/\omega_{TO}$) greater than 0.5, originating from the interaction of the ENZ mode with SiO$_2$ phonons. We also presented a theoretical framework (both classical and quantum mechanical) to model USC phenomena in the presence of the forbidden Reststrahlen band, which can be applied to homogeneous materials, waveguides, and resonant cavities. Our coaxial nanocavity platform features ultratight confinement of long-wavelength radiation while exhibiting extraordinary optical transmission (up to 60\% absolute transmission). 

The ENZ nanocavities can be filled with a wide range of metal oxides during the ALD step of the fabrication process (e.g. Al$_2$O$_3$, TiO$_2$, HfO$_2$, ZnO). Alternatively, the oxide can be selectively etched after fabrication, and different vibrationally-active materials can be incorporated into the gaps, thus enabling USC to be realized using a broad range of materials in the MIR, far-IR, and terahertz spectral range. 
Furthermore, since single-nanometer-thick phononic films separate the core and cladding electrodes in each coax, the platform provides a route to combine electron tunneling with vibrational USC and extraordinary optical transmission, which should enable observation of ground-state electroluminescence and the dynamic Casimir effect, and which should lead to the development of novel MIR optoelectronic devices. The ability to reach the USC regime in mass-produced nanocavity systems can also open up new avenues to explore non-perturbatively coupled light-matter systems, multiphoton effects, and higher-order nonlinear effects,  which may lead to novel applications in sensing, spectroscopy, nanocavity optomechanics, and chip-scale sources\cite{brongersma2010case}.


{\bf Acknowledgments}. D.Y., I.H.L., and S.-H.O. acknowledge support from the National Science Foundation (ECCS 1610333 and ECCS 1809723). F.dL.P. and L.M.M. acknowledge financial support from Spanish Ministry of Economy and Competitivity through projects MAT2017-88358-C3-1-R and MAT2017-88358-C3-2-R and the Aragón Government project Q-MAD. J.D.C. was supported by Office of Naval Research Grant N00014-18-12107.

{\bf Author contributions}. D.Y. performed device design, fabrication, and measurements. F.dL.P. and L.M.M. developed theories and performed numerical calculations. D.A.M. and I.-H.L. performed computer simulations. M.P., M.B.R., and J.D.C. analyzed the results. All authors analyzed the data and wrote the paper together.

{\bf Competing interests}: The authors declare no competing interests

{\bf Data availability}: The data that support the plots within this paper and other findings of this study are available from the corresponding author upon reasonable request.

\newpage
\appendix
\renewcommand{\thesection}{S\arabic{section}}
\renewcommand{\thetable}{S\arabic{table}}  
\renewcommand{\thefigure}{S\arabic{figure}}
\renewcommand{\theequation}{S\arabic{equation}}
\setcounter{figure}{0}
\setcounter{equation}{0}
\setcounter{table}{0}
\section*{Supporting Information}
\section*{Materials and methods}
 After pre-cleaning and pre-baking (200 $^\circ$ C for 3 min) processes, an undoped double-side-polished 4-inch (100) Si wafer (University wafer) was spin-coated with AZ MIR701 photoresist for 45 s at 5,000 r.p.m.then baked at 90 $^\circ$C for 90 s. The patterns of hexagonal hole arrays with 24 different diameters from 430 nm to 1120 nm were transferred onto a Si wafer substrate by photolithography (Canon 2500 i3 stepper) with a dose of 150 mJ/cm$^2$, followed by post-exposure bake at 110 $^\circ$ and development process with AZ 3000 MIF for 60 s. Hexagonal Au disk arrays were created on the Si substrate via directional evaporation of 3 nm Ti and 150 nm Au films (CHA, SEC 600), followed by a liftoff process with AZ kwik strip. After oxygen plasma cleaning (STS, 320PC) with 100 W for 30 s to remove photoresist residues, the Au disk arrays were coated conformally with a thin SiO$_2$ film using ALD (Cambridge Nano Tech Inc., Savannah) at a typical deposition rate of 1.2 $\AA$ per cycle, which was performed using ozone precursor and water vapor at 180 $^\circ$C. After conformal sputtering of 3 nm Ti and 400 nm Au (AJA, ATC 2200), the top surface of 400 nm-thick Au-deposited disk patterns was planarized by a glancing-angle ion milling using a 240 mA Ar ion beam incident at a 5 $^\circ$ from the horizontal plane (Intlvac, Nanoquest) until the top entrance of coaxial nanocavities is exposed.

\section*{Supplementary Text}
\section{Transmission resonances of a hole array}
\label{sec:TE}
Waveguide modes are the most natural choice for expanding the electromagnetic (EM) fields inside the coaxial holes \cite{FJRMP10}. In numerical simulations, convergence is fast achieved with a small number of such modes. In fact, considering just the fundamental mode is an excellent approximation for deeply subwavelength holes. 
The bouncing back and forth of the EM fields inside the holes leads to Fabry-Perot (FP) like transmission resonances. These resonances appear spectrally close to the cutoff condition of the fundamental waveguide mode inside the hole \cite{FJRMP10}, when it acts as an epsilon-near-zero (ENZ) medium.\cite{AluPRB08} When holes are arranged periodically, additional transmission resonances are developed at wavelengths slightly redrafted with respect to the array periodicity ($P$), due to the excitation of surface modes.  
A strong hybridization of these two kind of transmission resonances occurs when the cutoff wavelength of a FP resonance is near $P$.
However, our coaxial arrays are designed so that  FP transmission resonances appear in the mid IR while surface resonances are at the near IR, so that FP and surface modes do not hybridize. 

Fig. \ref{fig:cutoff}  compares the transmittance of the hole array with the square effective index $q^2_z=c^2 k^2_z/\omega^2$. Notice that Im$[k^2_z]$ does not vanish at cutoff due to the absorption losses. Even for uncoupled cavity modes, when the hole is filled with a lossless medium with dielectric constant $\epsilon_\infty$,  losses in the metal cause that Im$[k^2_z] \neq 0$, see magenta line in Fig. \ref{fig:cutoff}. 

Finally, it is worth to notice that, in the structures we consider, the gap 
size (between $G=2$ nm and $G=21$ nm)  is smaller than the skin depth ($\sim30$ nm). In these instances, the PEC approximation fails and both the field profile and the dispersion relation strongly depend on the penetration inside the metal, which thus must be taken into account in the calculation. Then, the fundamental waveguide mode has no longer a pure TM or TE character (as it occurs for a PEC), being instead a linear combination of the two polarizations. However, for the process of transmission and the range of frequencies considered, we find that the TE component couples to the external illumination much better than the TM one, so the mode behaves effectively as a quasi-TE mode. 

\section{Comparison of simulations and experiments}
\label{sec:compsimexp}

To further investigate the coupling between ENZ and SiO$_2$ phonon modes, we compare FEM numerical modeling (dispersion map) and experimental results (white dots) in Fig. \ref{fig:Tcountourplot}. FEM-modeled dispersion agrees well with the measured data, and we observe an anti-crossing band that is about two times wider than the Reststrahlen band. White circular dots are extracted from the resonance peaks at the higher-frequency-side. White square dots trace the resonance peaks at the lower-frequency-side, below the TO phonon resonance of SiO$_2$. This mode splitting in excess of the Reststrahlen bandwidth results in an unusually large forbidden energy gap compared with other strong coupling experiments involving phonon polaritons and MIR plasmonic resonators.

\section{Lattice model for a homogeneous medium}
\label{sec:bulklattice}
For completeness, this section summarizes the theory developed by Born and Huang [36] on the optical phonons in a diatomic crystal. In this case, there are three optical phonon branches (one longitudinal and two transversal). Unit cells with a larger number of atoms ($s$), and correspondingly $3(s-1)$ optical branches, will be discussed in Sec. \ref{sec:multphonons}.  
For polarizable (nonrigid) ions in the diatomic cell, the equation of motion and the polarization are given by 
\begin{eqnarray} 
\label{eq:xbulk}
\ddot{\mathbf{x}} &=& \gamma_{11} \mathbf{x}-\gamma_{13} \dot{\mathbf{x}}+\gamma_{12} \mathbf{E}, \\
\label{eq:Pbulk}
\mathbf{P} &=& \gamma_{12} \mathbf{x}+\gamma_{22} \mathbf{E}, 
\end{eqnarray}
where the normalized coordinate $\mathbf{x}=(\mathbf{u}_+-\mathbf{u}_-)/\sqrt{\mu n}$  depends on  the relative displacement of the positive and negative ions,  $\mathbf{u}_+-\mathbf{u}_-$, the reduced mass, $\mu=m_+ m_-/(m_+ + m_-)$, and the number of cells per unit volume, $n$. $\mathbf{P}$ and $\mathbf{E}$ are the dielectric polarization and electric field, respectively. By considering harmonic solutions, where the time dependence of all fields goes as $\mathbf{x}=\mathbf{x}_0 \exp(-i \omega t)$, we can eliminate $\mathbf{x}=-\gamma_{12}\mathbf{E}/(\omega^2+\gamma_{11}+i\gamma_{13} \omega)$ from these equations, and relate $\mathbf{P}$ and $\mathbf{E}$ by the equation $\mathbf{P}=[\gamma_{22}-\gamma^2_{12}/(\omega^2+\gamma_{11}+i\gamma_{13} \omega)]\mathbf{E}$. From the definition of the dielectric displacement, $\mathbf{D}=\mathbf{E}+4\pi \mathbf{P}=\epsilon \mathbf{E}$, we obtain the dielectric function, $\epsilon(\omega)=1 +4 \pi \gamma_{22}-4 \pi \gamma^2_{12}/(\omega^2+\gamma_{11}+i\gamma_{13} \omega)$. If $\epsilon(\omega)$ is written in terms of measurable quantities,
\begin{eqnarray}
\epsilon(\omega)=\epsilon_\infty\left(1+\frac{\omega^2_p}{\omega^2_{TO}-\omega^2-i\gamma \omega} \right), \nonumber
\end{eqnarray}
the $\gamma$-coefficients can be expressed as $\gamma_{22}=(\epsilon_\infty-1)/4\pi$, $\gamma_{12}=(\epsilon_\infty/4\pi)^{1/2} \omega_p$, $\gamma_{13}=\gamma$, and $\gamma_{11}=-\omega^2_{TO}$.

Also notice that, neglecting the induced internal polarization in the ions (i.e., considering them as rigid), the vibrational coupling constant  $\omega_p$ can be estimated to be [36]
\begin{eqnarray}
\label{eq:wp}
\omega_p=\sqrt{\frac{4 \pi n e^2}{M}}. 
\end{eqnarray}

\section{Wave equation}
\label{sec:we}
The wave equation may be derived from Maxwell's equations (Gaussian convention),
\begin{eqnarray}
\label{eq:ampere}
\nabla \times \mathbf{H} &=&\frac{1}{c}\frac{\partial\mathbf{D}}{\partial t},  \\
\label{eq:faraday}
\nabla \times \mathbf{E} &=&- \frac{1}{c} \frac{\partial\mathbf{H}}{\partial t},
\end{eqnarray}
in the following way. After assuming  harmonic solutions ($\sim \exp(-i \omega t)$), we can apply the operator $\nabla \times$ on (\ref{eq:faraday})  and replace (\ref{eq:ampere}) into the resulting equation to have
\begin{eqnarray}
\label{eq:we}
\nabla \times \nabla \times \mathbf{E}  -\frac{\omega^2}{c^2} \mathbf{D}=0. 
\end{eqnarray}
Using the constitutive relation for the displacement current $\mathbf{D}=\mathbf{E}+4\pi \mathbf{P}=\epsilon \mathbf{E}$, we obtain Eqs.(1) and (4) of the main text. 

\section{Scaling relation}
\label{sec:scale}
At a given frequency $\omega$ and for a general dielectric constant $\epsilon(\mathbf{r})$, the wave equation for the Electric field (\ref{eq:we}),
\begin{equation}
\label{eq:maxwell}
\nabla \times \nabla \times \mathbf{E}(\mathbf{r}) - \epsilon(\mathbf{r}) \frac{\omega^2}{c^2} \mathbf{E}(\mathbf{r}) = 0, 
\end{equation}
only depends on the combination  $\kappa^2(\mathbf{r}) \equiv  \epsilon(\mathbf{r}) (\omega/c)^{2}$. For a waveguide defined by an internal dielectric region (characterized by   $\epsilon_{d}$) and an external metallic region (with $ \epsilon_{m}$), the field only depends on the values of the corresponding  $\kappa_{d}^2 $ and $\kappa_{m}^2$.  The scaling relation 
\begin{eqnarray}
\epsilon_d, \epsilon_m, \omega \rightarrow \epsilon_d \eta, \epsilon_m \eta, \omega/\sqrt{\eta} \nonumber
\end{eqnarray}
leaves $\kappa_{d}^2 $ and $\kappa_{m}^2$ unchanged for any $\eta$. However,  for a given metal defining the waveguide, if the material inside the waveguide is replaced by a different dielectric, the scaling relation will not be fulfilled, as the spectral dependence of $\epsilon_m$ will generally not follow the scaling imposed by the change in $\epsilon_d$.  Nevertheless, we find two relevant exceptions: a perfect electrical conductor and a Drude metal at low frequencies. 

In a PEC, $\epsilon_m= -\infty$. In that case, the dependence on $\epsilon_m$ disappears from the scaling relation (as $\epsilon_m \eta = -\infty$, independent of $\eta$). If Eq. (\ref{eq:maxwell}) holds and the waveguide is filled with $\epsilon'_d(\omega)$, the EM fields satisfy the same Maxwell's equation with $\kappa'^2_d = \epsilon'_d(\omega) (\omega/c)^{2}$ so that the solution $\mathbf{E}$ for $\epsilon_{d}$ at a given frequency $\omega$ is the same than for $\epsilon'_d(\omega')$ at a different $\omega'$ given by
\begin{eqnarray}
\label{eq:propcond}
\epsilon'_d(\omega') \omega'^2=\epsilon_{d} \omega^2. 
\end{eqnarray}
A similar behavior is found for a Drude metal with dielectric function $\epsilon_m(\omega)=1-\omega^2_{pm}/\omega^2$. At frequencies much smaller than the plasma frequency $\omega_{pm}$, i.e. at MIR and smaller frequencies, $\kappa_{m}^2 \approx - \omega_{pm}^{2}/c^2$ and does not depend on $\omega$. Therefore, field patterns only depend on $\omega_{pm}$ and $\kappa_{d}^2$.  These trends are observed for the spectral window considered in the experiments.

In particular, Eq. (\ref{eq:propcond}) is fulfilled when the field pattern corresponds to the cutoff condition $k^2_{z}=0$, where $k_z$ is the component of the wavevector along the waveguide $(z)$ axis. Thus, if we know the cutoff frequency $\omega_{c_\infty}$ when a waveguide is filled with a material characterized by $\epsilon_\infty$, we can obtained the cut-off frequency when the material filling the waveguide is characterized by a dielectric constant $\epsilon(\omega)$:  
\begin{equation}
\label{eq:cutoff}
\epsilon(\omega_{c}) \, \omega_{c}^{2} = \epsilon_\infty \, \omega_{c_\infty}^{2}. 
\end{equation}
Notice that in the case of the strong frequency dependence of some dielectrics in the region of anomalous dispersion, this equation may (and in the case of SC does) have more than one solution.

Eq.  (\ref{eq:cutoff}) remains an excellent approximation even for the experimental case of an array of annular holes drilled in a gold film, where holes are filled with SiO$_2$. Despite absorption losses in both Au and SiO$2$,  the right cutoff frequencies, $\omega_+$ and $\omega_-$, are predicted by Eq. (\ref{eq:cutoff}). These cutoff frequencies are represented with vertical dash-dotted lines in Fig. \ref{fig:cutoff}. Eq.  (\ref{eq:cutoff}) works well in this case for the lossy response of the metal is already included in $\omega_{c_\infty}$, which is used as an input parameter.

\section{Phonon polariton branches in a coaxial waveguide, for a single vibrational mode}
\label{sec:phonpol}
The interaction of optical vibrational modes and the electromagnetic field is particularly important when frequencies and wavevectors of phonon and photon fields coincide near the crossover of the corresponding dispersion relations. The resulting phonon-polariton fields are the simultaneous solutions of Maxwell (\ref{eq:ampere}, \ref{eq:faraday}) and lattice (\ref{eq:xbulk}, \ref{eq:Pbulk}) equations, which are obtained from the Lagrangian density,
\begin{eqnarray}
 \label{eq:Lden}
 \mathcal{L}=\frac{1}{8\pi} \left(\epsilon_\infty  \mathbf{E}^2- \mathbf{H}^2 \right)+\frac{1}{2} \left( \dot{\mathbf{x}}^2-\omega^2_{TO} \mathbf{x}^2 \right)+\sqrt{\frac{\epsilon_\infty}{4 \pi}} \omega_p \; \mathbf{x} \cdot \mathbf{\mathbf{E}},
 \end{eqnarray}
 using the Lagrangian equations.
In this section, we compute the polaritonic branches of an infinite coaxial waveguide assuming that light interacts with a single vibrational mode with frequency $\omega_{TO}$ and coupling constant $\omega_p$. The electronic background of the ions is characterized by the high frequency dielectric constant $\epsilon_\infty$.

For the sake of convenience, the magnetic field is eliminated from Maxwell's equations (\ref{eq:ampere}, \ref{eq:faraday}) and the resulting wave equation (\ref{eq:we}) is employed in what follows. As dispersive  terms of vibrational modes (proportional to the second derivative of the displacement field) are not included in Eq. (\ref{eq:Lden}), the dynamical lattice equations derived with the Lagrangian equations coincide with those for the bulk (Eqs. \ref{eq:xbulk}, \ref{eq:Pbulk}). The relative displacement field  of the ions, $\mathbf{x}$, and its polarization field, $\mathbf{P}$ resulting from lattice equations have the same symmetry as the electric field, $\mathbf{E}$, inside the waveguide and are parallel to it. This local relation between $\mathbf{x}$, $\mathbf{P}$, and $\mathbf{E}$ will be valid as long as the dispersion relation of the phonons is neglected, which is a good approximation because the $k$ dependence of the uncoupled optical phonons is negligible compared to that of photons.

Let us consider a given waveguide mode M, characterized by a wavevector $k$ along the waveguide axis  (for instance $M=TE_{11}$, the fundamental mode of the coaxial waveguide described in Sec. \ref{sec:TE}).  When the aperture is filled with a uniform dielectric constant $\epsilon_\infty$, the electric field satisfies the wave equation (\ref{eq:we})
\begin{eqnarray}
\nabla \times \nabla \times E \mathbf{E}_M   -\epsilon_\infty \frac{\omega^2_k}{c^2} E \mathbf{E}_M =0, \nonumber
\end{eqnarray}
where $\mathbf{E}=E\mathbf{E}_M$ is electric-field vector with amplitude  $E$ and $\mathbf{E}_M$ is the normalized transverse solution of Maxwell equations for mode M at frequency $\omega_k$. The wave equation transforms to 
\begin{eqnarray}
\nabla \times \nabla \times E \mathbf{E}_M  - \frac{\omega^2}{c^2} \left( E+4\pi P\right) \mathbf{E}_M=0. \nonumber
\end{eqnarray}
in order to accommodate the polarization field, $\mathbf{P}=P\mathbf{E}_M$, of the oscillating ions when the waveguide is filled with the phononic material. $\mathbf{E}_M$ satisfies both wave equations if the scaling relation (\ref{eq:propcond}) is fulfilled, i.e.
\begin{eqnarray}
\label{eq:Eprop}
\omega^2(E+4\pi P)=\omega_k^2 E.
\end{eqnarray}
Sec. \ref{sec:scale} shows it is a good approximation for the coaxial hole array studied experimentally.

Both the relative displacement field and the polarization field for vibrational modes without spatial dispersion are determined by the given mode profile of the electric field, i.e. $\mathbf{x}(\mathbf{r})=x \mathbf{E}_M(\mathbf{r})$ and $\mathbf{P}(\mathbf{r})=P \mathbf{E}_M(\mathbf{r})$, and, therefore,  lattice equation (\ref{eq:xbulk}, \ref{eq:Pbulk}) may by written in terms of the field amplitudes $x$, $P$, and $E$,
\begin{eqnarray}
\label{eq:newton}
-\omega^2 x &=& -\omega^2_{TO} x+i \gamma \omega x+\sqrt{\frac{\epsilon_\infty}{4\pi}} \omega_p E, \\
\label{eq:pol}
P &=& \sqrt{\frac{\epsilon_\infty}{4\pi}} \omega_p x+ \frac{\epsilon_\infty-1}{4\pi} E,
\end{eqnarray}
where a phenomenological damping force, $-\gamma \dot{x}$ , has been added.  
Eliminating $P$ from equations (\ref{eq:Eprop}-\ref{eq:pol}), the system of equations can be expressed in a matrix form, stating that $\mathbf{E}_M (\mathbf{r},\mathbf{k})$  is still a solution of Maxwell equations but at a frequency $\omega$ satisfying
\begin{eqnarray}
\left( \begin{array}{cc}
\omega^2-\omega^2_{TO}+i \gamma \; \omega &  \omega \; \omega_p  \\
\omega \; \omega_p &  \omega^2 -\omega^2_k+i\delta \; \omega_k
\end{array} \right) \cdot \left( \begin{array}{c}
\omega \; x \\ \sqrt{\epsilon_\infty/4\pi}  E
\end{array} \right) &=& 0. \nonumber 
\end{eqnarray}  
We have included a finite linewidth in the photon mode by making the replacement $\omega_k \rightarrow \omega_k-i \delta/2$. 

In the strong coupling regime, one can assume that  $\omega_{TO} \gg \omega_p \gg \gamma, \delta$ at the crossing point of the photon and TO phonon frequencies $(\omega_k=\omega_{TO})$, we find from the secular equation (setting the determinant of the matrix equal to zero) that the degeneracy between the two modes is lifted,
\begin{eqnarray}
\omega_\pm=\omega_{TO} \pm \frac{\omega_p}{2}-i\frac{\gamma_{FP}+\gamma}{4}, \nonumber
\end{eqnarray}
and the frequency splitting, $\Delta \omega=\omega_+-\omega_-=\omega_p$, coincides with $\omega_p$ in  this simple case of one phonon mode. Moreover, the width of each dressed state is the average of the photon ($\delta/2$) and phonon ($\gamma/2$) linewidths. The splitting can be resolved only if $\Delta \omega > (\gamma+\delta)/2$. 

 In the USC regime, however, $\omega_p$ is not longer a small parameter and we need to compute the exact values of the polaritonic frequencies at resonance,
\begin{eqnarray}
\omega_\pm=\sqrt{\omega^2_{TO}+\frac{\omega^2_p}{4}}\pm \frac{\omega_p}{2}, \nonumber
\end{eqnarray}
where absorption have been neglected for the sake of simplicity. 
We obtain the same splitting though the central frequency is blue-shifted from $\omega_{TO}$ to $\sqrt{\omega^2_{TO}+\omega^2_p/4}$. Therefore, an appreciable blue shifting of the central frequency is a signature of the USC regime. This blue shift is apparent in the experimental results reported in  Fig. \ref{fig:fit}(b).  We find a blue shift of up to 11\% of $\omega_{TO}$ (see Table 1 in the main text). This behavior cannot be explained by the constant and smaller value of 3\% provided by the model that takes into account a single vibrational model of SiO$_2$ (see Fig. S6(b)). Rather, experimental trends are  reproduced only when several vibrational modes are taken into account (see Sec. S8). 

Additional remarks:
\begin{itemize}
\item The frequency dependence of the off-diagonal coupling term is related to the long-range nature of the Coulomb forces (i.e the dipole-dipole infraction induced via an electric field), and does not appear in the text-book example of two harmonic oscillators (Sec. \ref{sec:cho}), due to the short-range nature of the oscillator restoring forces.
\item The Reststrahlen band is smaller than the splitting, i.e $\omega_{L0}-\omega_{TO}=(\omega_p/2\omega_{TO}) \omega_p<\Delta \omega$, and thus, the splitting is not simply the manifestation of absorption induced within the Reststrahlen band.  
\item The asymptotic limits of the general solution are
\begin{eqnarray}
\lim_{\omega'_k \rightarrow 0} \omega_-&=& \omega_k \sqrt{\epsilon_\infty/\epsilon_0}, \;\;\;\; \mbox{photon-like} \nonumber \\
\lim_{\omega'_k \rightarrow 0} \omega_+&=& \omega_{LO}, \;\;\;\;\;\;\;\;\;\;\;\;\;\; \mbox{phonon-like} \nonumber \\
\lim_{\omega'_k \rightarrow \infty} \omega_-&=& \omega_{TO}, \;\;\;\;\;\;\;\;\;\;\;\;\;\; \mbox{phonon-like} \nonumber \\
\lim_{\omega'_k \rightarrow \infty} \omega_+&=& \omega_k. \;\;\;\;\;\;\;\;\;\;\;\;\;\;\;\;\; \mbox{photon-like} \nonumber
\end{eqnarray}
The existence of 4 different asymptotes is a clear difference with respect to the case of two harmonic oscillators coupled with short-range interactions, which present only three asymptotic values (as the asymptotes represented by $\omega_{TO}$ and $\omega_{LO}$ would coincide in this case).

\item  The vibrational coupling constant is a function of the molecular concentration $n$, $\omega_p=\sqrt{4 \pi n e^2/M}$ (Eq. \ref{eq:wp}).  The dependence  $\omega_p\sim \sqrt{n}$ is shared by classical, semi-classical and quantum descriptions \cite{TormaRPP15}. This behavior has been confirmed experimentally for a classical Fabry-Perot  resonator \cite{ZhuPRL90}.
\end{itemize}

\section{Coupled harmonic oscillators} 
\label{sec:cho}
\begin{figure}[H]
\centering
\includegraphics[scale=0.4]{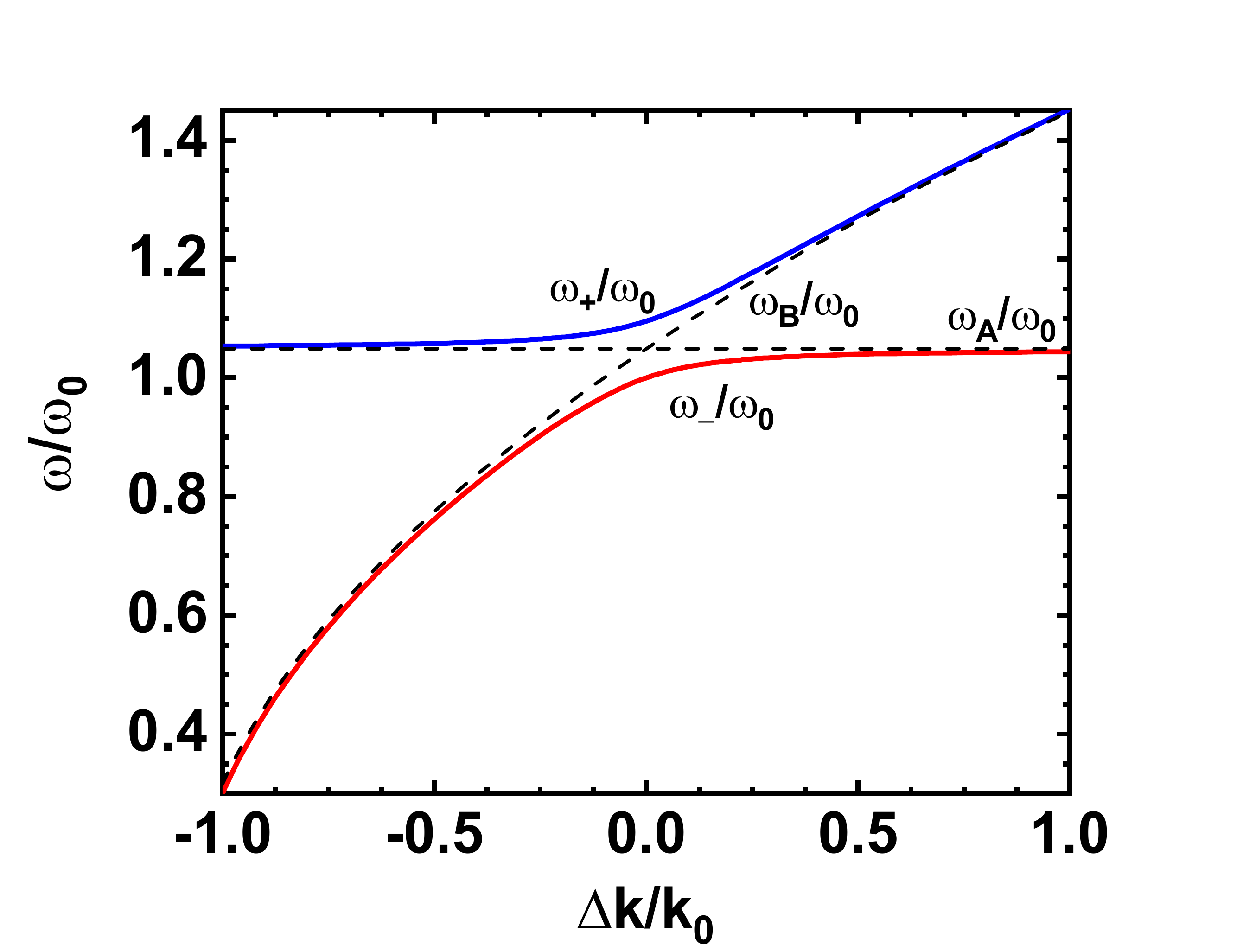}
\caption{Bare (dashed lines) and dressed (solid lines) frequencies  for a harmonic oscillator with $\omega_A=\omega_0 \sqrt{1+\kappa/k_0}$, $\omega_B=\omega_0 \sqrt{1+\Delta k/k_0+\kappa/k_0}$, and $\kappa=0.1 \; k_0$. }
\label{fig:osc}
\end{figure}

We consider two mechanical oscillators with eigenfrequencies $\sqrt{k_A/m_A}$ and $\sqrt{k_B/m_B}$ coupled by a spring with constant $\kappa$ as a canonical example for strong coupling \cite{Novotny}. The differential equations 
\begin{eqnarray}
m_A \ddot{x}_A+k_A x_A+\kappa \left( x_A-x_B\right)&=& 0, \nonumber \\
m_B \ddot{x}_B+k_B x_B-\kappa \left( x_A-x_B\right)&=& 0, \nonumber 
\end{eqnarray} 
describe the motion of the system. The harmonic solutions of these equations can be written in a matrix form
\begin{eqnarray}
\left( \begin{array}{cc}
\omega^2 -\omega^2_A &  \Gamma \\
\Gamma & \omega^2-\omega^2_B
\end{array} \right) \cdot \left( \begin{array}{c}
\sqrt{m_A} x_A \\ \sqrt{m_B} x_B
\end{array} \right) &=& 0, \nonumber 
\end{eqnarray} 
where  $\omega^2_A=(k_A+\kappa)/m_A$, $\omega^2_B=(k_B+\kappa)/m_B$, and $\Gamma=\kappa/\sqrt{m_A m_B}$. 
The diagonal terms  contain the eigenfrequencies of the uncoupled oscillators, modified by the interaction, while the off-diagonal terms are proportional to the coupling strength. The secular equation yields the dressed frequencies of the system. 
\begin{eqnarray}
\omega^2_\pm=\frac{\omega^2_A+\omega^2_B \pm \sqrt{(\omega^2_A-\omega^2_B) ^2+4 \Gamma^2}}{2}. \nonumber
\end{eqnarray}
The frequency splitting, $\omega_+-\omega_- \approx \Gamma$, increases with the coupling constant. A numerical example is presented in Fig. \ref{fig:osc}.

\section{Multiple vibrational degrees of freedom}
\label{sec:multphonons}
Sec. \ref{sec:phonpol} discusses polaritonic branches in a coaxial waveguide filled with a material with a lossless single vibrational optical mode. We consider now the influence of additional vibrational modes, with damping, in the optical response of the filled coaxial aperture. The lattice equations of motion for a unit cell with $N$ oscillator normal modes  read
\begin{eqnarray}
\ddot{x}_1 &=& -\omega^2_{TO_1} x_1-\gamma_1 \dot{x}_1  +\sqrt{\frac{\epsilon_\infty}{4 \pi}} \omega_{p_1} E, \nonumber \\
\ddot{x}_2 &=& -\omega^2_{TO_2} x_2-\gamma_2 \dot{x}_2 +\sqrt{\frac{\epsilon_\infty}{4 \pi}} \omega_{p_2} E, \nonumber \\
&\vdots& \nonumber \\
\ddot{x}_N &=& -\omega^2_{TO_N} x_N-\gamma_N \dot{x}_N +\sqrt{\frac{\epsilon_\infty}{4 \pi}} \omega_{p_N} E, \nonumber \\
P &=& \sqrt{\frac{\epsilon_\infty}{4 \pi}} \sum^N_{i=1}\omega_{p_i} x_i+\frac{\epsilon_\infty-1}{4 \pi} E. \nonumber
\end{eqnarray}
Simultaneous solution of lattice and Maxwell's equations for transverse modes can be expressed as a  matrix equation of order $N+1$,
\begin{eqnarray}
\left( \begin{array}{ccccc}
 \omega^2-\omega^2_{TO_1}+ i \gamma_1 \omega  & 0 &  \dots  & 0 &  \omega \omega_{p_1} \\
0 & \omega^2-\omega^2_{TO_2}+ i \gamma_2 \omega  &  \dots  & 0 &  \omega \omega_{p_2} \\
\vdots & \vdots & \vdots & \vdots & \vdots \\
0 & 0  &  \dots  & \omega^2-\omega^2_{TO_N}+ i \gamma_N \omega &  \omega \omega_{p_N} \\
\omega \omega_{p_1}  & \omega \omega_{p_2} & \cdots  &  \omega \omega_{p_N} & \omega^2 -\omega'^2_k
\end{array} \right) \cdot \mathbf{F} &=& 0, \nonumber \\
\label{eq:multimatrix}
\end{eqnarray}
with $\mathbf{F}=(\omega x_1,\omega x_2,\dots,\omega x_N,\sqrt{\epsilon_\infty/4\pi} E)^t$.
The secular equation for this matrix can be written as
\begin{eqnarray}
\epsilon(\omega) \omega^2=\omega^2_k, \nonumber
\end{eqnarray}
provided that the effective dielectric function is generalized to include all vibrational modes,
\begin{eqnarray}
\label{eq:epswmulti}
\epsilon(\omega)=\epsilon_\infty \left( 1+\sum^N_{j=1} \frac{\omega^2_{p_j}}{\omega^2_{TO_j}-\omega^2-i \gamma_j \omega }\right). 
\end{eqnarray}
In Fig. \ref{fig:fit}(a), the experimental spectral position of transmission resonances for an array of coaxial holes filled with SiO$_2$, are compared with those calculated with the formalism described above. Calculations were done for two models for the dielectric constant of SiO$_2$: in one of them we considered that only one vibrational mode contributed to $\epsilon_{SiO_2}$, while the other considered 3 vibrational modes (the contribution from each vibrational mode to  $\epsilon_{SiO_2}$ is described by a Lorentzian function using Palik’s data \cite{Palik}. For a single vibrational mode we find a good agreement with the upper band but the lower band is poorly described. The agreement with both bands is improved when two lower-frequency vibrational modes are added. Notice that the size of the main gap is reduced in this case. 

Fig. \ref{fig:fit}(b) shows the blue shifting, $\Delta \omega_{cen}=(\omega_{cen}-\omega_{TO_1})/\omega_{TO_1}$ $[\times 100\%]$, of the central frequency, $\omega_{cen}=(\omega_++\omega_-)/2$, computed  from the experimental values of the upper ($\omega_+$) and lower ($\omega_-$) polaritonic frequencies at resonance as a function of the experimental gap sizes, $G=2$, 7, 14, and 21 nm. $\Delta \omega_{cen}$ increases from 6\% at $G=2$ nm to 11\% at $G=21$ nm. This behavior cannot be explained by the constant value of 3\% predicted by our model if we assume that a single vibrational model of SiO$_2$ is interacting with the cavity photons. A better agreement is obtained when 3 vibrational modes are taken into account, see Fig. \ref{fig:fit}(b). Therefore, we conclude that the interaction with several vibrational modes reduces the gap size of the main gap but increases the blue shifting of its central frequency. 

\section{Quantum theory of phonon polaritons}
\label{sec:QP}
In this section, we present the quantum-mechanical treatment of the coupled photon and phonon modes in a coaxial cavity of length $L$ with ideal PEC walls. Such simplified boundary conditions cannot provide a good quantitative agreement with the polaritonic frequencies experimentally measured for gap sizes smaller than the skin depth. However, the qualitative results derived here will validate the classical approach of Sec. \ref{sec:phonpol}.    

We will follow the canonical procedure for second quantization and write a Hopfield-like Hamiltonian in the representation of the number operators for photons and phonons. We start from the classical Lagrangian of the system under study, which is the integral of the Lagrangian density (Eq. \ref{eq:Lden}) previously used for computing the classical equations of motion , 
 \begin{eqnarray}
 L=\int \left[ \frac{1}{8\pi} \left( \frac{\epsilon_\infty }{c^2} \dot{\mathbf{A}}^2-\left(\nabla \times \mathbf{A}\right)^2 \right)+\frac{1}{2} \left( \dot{\mathbf{x}}^2-\omega^2_{TO} \mathbf{x}^2 \right)-\frac{\gamma_{12}}{c}\mathbf{x} \cdot \mathbf{\dot{\mathbf{A}}} \right] d^3r,  \nonumber
 \end{eqnarray}
where, for the sake of convenience, electric and magnetic fields have been expressed as a function of the vector potential $\mathbf{A}$ (in the Coulomb gauge, $\nabla \cdot \mathbf{A}=0$), i.e. $\mathbf{E} = \dot{\mathbf{A}}/c$, $\mathbf{H} = \nabla \times \mathbf{A}$.
$\mathbf{A}=A \mathbf{A}_m(\mathbf{r})$ is a solution of Maxwell's wave equations with the appropriate boundary conditions for a given mode $m$, $A$ is the field amplitude, and $\mathbf{A}_m(\mathbf{r})$ the mode profile. The mode is defined as $m=\{M,k_z\}$, where $M$ takes into account both polarization and field profile in the $XY$ plane and, along the direction perpendicular  to the plane, $k_z=\pi \ell/L$, with $\ell=1$,2,3,$\dots$. In the numerical calculations below we will only consider the fundamental $M=$TE$_{11}$ mode for a given $k_z$.  As we assume a non-dispersive vibrational mode with frequency $\omega_{TO}$, the relative displacement field  is proportional to the vector potential inside the cavity, $\mathbf{x}=x \mathbf{A}_m(\mathbf{r})$. 
$\gamma_{12}^2=\epsilon_\infty \omega_p^2/4 \pi$ is the classical coupling constant.

Following Hopfield \cite{HopfieldPR58}, the Lagrangian  $L$ can be transformed to
\begin{eqnarray}
L' &=& L+\int \frac{\gamma_{12}}{c}\frac{\partial \left( \mathbf{x} \cdot \mathbf{A}\right)}{\partial t}d^3r. \nonumber 
 \end{eqnarray}
The Hamiltonian of the system is derived from the Lagrangian function with help of the canonical momentum fields, $\mathbf{B}=\partial \mathcal{L}/\partial \dot{\mathbf{A}}=\epsilon_\infty \dot{\mathbf{A}}/4 \pi c^2$ and  $\mathbf{Y}=\partial \mathcal{L}/\partial \dot{\mathbf{x}}=\dot{\mathbf{x}}+\gamma_{12} \mathbf{A}/c$ and the canonically conjugate operators,
\begin{eqnarray}
\hat{\mathbf{A}} &=& \sum_m \sqrt{\frac{2 \; \pi \hbar \; c^2}{V_m  \; \omega_m \; \epsilon_\infty}} \; \mathbf{A}_m(\mathbf{r}) \left(a_m +a^+_m  \right),  \nonumber \\
\hat{\mathbf{B}} &=& i \sum_m \sqrt{\frac{\hbar  \; \omega_m \; \epsilon_\infty}{8 \; \pi c^2 V_m}} \; \mathbf{A}_m(\mathbf{r}) \left(a^+_m  -a_m  \right),  \nonumber \\
\hat{\mathbf{x}} &=& \sum_m \sqrt{\frac{\hbar}{2 \; V_m  \; \omega_{TO}}} \; \mathbf{A}_m(\mathbf{r}) \left(b_m+b^+_m  \right),  \nonumber \\
\hat{\mathbf{Y}} &=& i \sum_m \sqrt{\frac{\hbar \; \omega_{TO}}{2 \; V_m }} \; \mathbf{A}_m(\mathbf{r}) \left(b^+_m -b_m  \right), \nonumber
\end{eqnarray}
where $\omega_m$ is the photon frequency, $a^+$ ($a$) and $b^+$ ($b$) are the creation (annihilation) operators for phonons and photons, respectively, and $V_m$ is the mode volume for both fields. 
These operators produce the Hamiltonian in the new representation, $H=H_{photon}+H_{phonon}+H_{int}$, with
\begin{eqnarray}
H_{photon} &=& \sum_m \hbar \omega_m \left( a^+_m a_m+\frac{1}{2}\right), \nonumber \\
H_{phonon} &=& \sum_m \hbar \omega_{TO} \left( b^+_m b_m+\frac{1}{2}\right), \nonumber \\
H_{int} &=& \sum_m \hbar \left[i C_m \left(a^+_m b_m-a_m  b^+_m \right)+D_m  \left(2a^+_k a_m+1 \right) \right. \nonumber \\
& & \left. +iC_m\left(a_m b_m -a^+_m b^+_m \right) +D_m  \left(a_m a_m+a^+_m a^+_m \right)  \right]. \nonumber
\end{eqnarray}
$H_{photon}$ and $H_{phonon}$ describe the energy of bare cavity photons and phonons in terms of their respective number operators $a^+_m a_m$ and $b^+_m b_m$. Several terms contribute to $H_{int}$ that depends on the coupling constants
\begin{eqnarray}
C_m &=& \frac{\omega_p}{2} \sqrt{\frac{\omega_{TO}}{\omega_m}}, \;\;
D_m = \frac{\omega^2_p}{4 \omega_m}. \nonumber
\end{eqnarray}
Terms in the first line constitute the resonant part of the light-matter interaction \cite{CiutiPRB05}. The term proportional to $C_m$ describes the creation (annihilation) of one photon and the annihilation (creation) of a phonon with the same wave number $m$. The term proportional to $D_m$ comes from the $\mathbf{A}^2$ term in the original Hamiltonian. It contains the photon number operator that produces a blue shifting (as $D_m>0$) of the bare cavity phonon frequency. In the second line of $H_{int}$, we find the antiresonant terms that are frequently neglected when the Hamiltonian is diagonalized. These terms produce the simultaneous destruction or creation of two excitations. 

We will diagonalize the full Hamiltonian (including the antiresonant terms) using the  Bogoliubov transformation,
\begin{eqnarray}
p_{mi}=w_{mi} a_m+x_{mi} b_m+y_{mi} a^+_m+z_{mi} b^+_m. \nonumber
\end{eqnarray}
The Bose commutation rule $ \left[ p_{mi},p^+_{m'i'}\right]=\delta_{i,i'} \delta_{m,m'}$,  where $i$, $i'$ are two solutions of the secular equation, 
imposes the normalization condition, $w^*_{mi} w_{mi'}+x^*_{mi} x_{mi'}-y^*_{mi} y_{mi'}-z^*_{mi} z_{mi'}=\delta_{i,i'}$, to the Hopfield coefficients. From the equation of motion for operators in the Heisenberg picture,
\begin{eqnarray}
\left[ p_{mi},\hat{H} \right]= \hbar \omega \; p_{mi}, \nonumber
\end{eqnarray}
the eigenvalue problem may be written in a matrix form,
\begin{eqnarray}
\left(\begin{array}{cccc}
\omega_m+2D_m-\omega & -i C_m & -2D_m & -i C_m \\
i C_m & \omega_{TO}-\omega & -i C_m & 0 \\
2D_m & -i C_m & -\omega_m-2D_m-\omega & -i C_m \\
-i C_m & 0 & i C_m & -\omega_{TO}-\omega
\end{array} \right) \left( \begin{array}{c}
w_{mi} \\ x_{mi} \\ y_{mi} \\ z_{mi}
\end{array} \right)=0. \nonumber
\end{eqnarray}
The determinant of the matrix, $\omega^4-\left(\omega^2_m +\omega^2_{TO}+\omega^2_p \right) \omega^2+\omega^2_m \omega^2_{TO}=0$, is identical to the one obtained with a classical approach (if absorption is neglected), i.e. quantum and classical descriptions provide the same polaritonic branches.. Therefore, this equation may be rewritten in the form $\epsilon(\omega) \omega^2=\epsilon_\infty \omega^2_m$, 
where $\epsilon(\omega)=\epsilon_\infty \left[1+\omega^2_p/(\omega^2_{TO}-\omega^2)\right]$. 

In Fig.  \ref{fig:ZP1}(a) the measured polaritonic frequencies are fitted with our model assuming that the fundamental cavity photon is interacting with the TO phonon. The  corresponding Hopfield coefficients $|w-y|^2$ and $|x-z|^2$, which
are solutions of the matrix equation for a given eigenvalue, are depicted in Fig. \ref{fig:ZP1}(b). The Hopfield coefficients evaluate the relative contribution of photon and phonon modes to the polaritonic states. 

The quantum approach accounts for the content of bare modes in the ground state energy. The  ground state of polariton excitations, $\left. \vert G \right\rangle$, contains a finite number of virtual cavity photons and phonons per mode \cite{CiutiPRB05},
\begin{eqnarray}
\left\langle G \vert a^+_m a_m \vert G \right\rangle &=& \sum_{i=\pm} \vert y_{im} \vert^2, \nonumber \\
\left\langle G \vert b^+_m b_m \vert G \right\rangle &=& \sum_{i=\pm} \vert z_{im} \vert^2, \nonumber 
\end{eqnarray}
where  $i=\pm$ for the two polaritonic branches. The  difference between the ground state energy of the polaritons, $E_G$, and ground state energy of the uncoupled system, $E_0$ , per mode 
\begin{eqnarray}
E_G-E_0=\hbar \Delta \omega_{ZP}= \hbar D_m-\sum_{i=\pm} \hbar \omega_{i} (\vert y_{im} \vert^2,+\vert z_{im} \vert^2), \nonumber 
\end{eqnarray}
is a function of the number of virtual modes \cite{QuattropaniNC86}. We find in the first term of this expression the positive coupling constant $D_m$ and in the second term the negative correlation contribution for a given mode, $\hbar \Delta \omega^{cor}_{ZP}$ for the two polaritonic frequencies $\omega_\pm$ \cite{CiutiPRB05}. These quantities can be written as a function of the coupling strength $\eta=\omega_p/\omega_{TO}$ at resonance ($\omega_m=\omega_{TO}$) if we use the exact value of the polaritonic frequencies, $\omega_\pm / \omega_{TO}=\sqrt{1+\eta^2/4} \pm \eta/2$, and retain the leading terms in the Series expansion for both the positive coupling constant $D_m$ and the Hopfield coefficients,
\begin{eqnarray}
\frac{D_m}{\omega_{TO}} &=& \frac{\eta^2}{4}, \nonumber \\
\frac{\Delta \omega^{cor}_{ZP}}{\omega_{TO}} &=& -\frac{\eta^2}{8}  \sqrt{1+\frac{\eta^2}{4}}. \nonumber
\end{eqnarray}
Calculated values of these quantities and the resulting zero-point differential frequency, $\Delta \omega_{ZP}=D_m-\Delta \omega^{cor}_{ZP} $ are reported in Table \ref{table:ZOE} for measured values of $\eta$. We find that $\Delta \omega_{ZP}$ is about a 3\% of the resonance frequency $\omega_{TO}$. Moreover, Fig. \ref{fig:ZP1} (a) compares the polaritonic branches with the exact differential zero point frequency $\Delta \omega_{ZP}$, which increases from 2\% of $\omega_{TO}$ at diameter $D=460$ nm to 5\% of $\omega_{TO}$ at $D=1120$ nm.

\section{Hopfield-like Hamiltonian for multiple vibrational degrees of freedom}
\label{sec:QPmulti}
The Hopfield Hamiltonian can be easily generalized to include $N$ vibrational modes,
\begin{eqnarray}
H_{photon} &=& \sum_m \hbar \omega_m \left( a^+_m a_m+\frac{1}{2}\right), \nonumber \\
H_{phonon} &=& \sum_{jm} \hbar \omega_{TO_j} \left( b^+_{jm} b_{jm}+\frac{1}{2}\right), \nonumber \\
H_{int} &=& \sum_{jm} \hbar \left[i C_{jm} \left(a^+_m b_{jm}-a_m  b^+_{jm} \right)+D_{jm}  \left(2a^+_k a_m+1 \right) \right. \nonumber \\
& & \left. +iC_{jm}\left(a_m b_{jm} -a^+_m b^+_{jm} \right) +D_{jm}  \left(a_m a_m+a^+_m a^+_m \right)  \right], \nonumber
\end{eqnarray}
with $j=1,\dots,N$. A new Bogoliubov transformation,
\begin{eqnarray}
p_{mi}=w_{mi} a_m+\sum_j x_{jmi} b_{jm}+y_{mi} a^+_m+\sum_jz_{jmi} b^+_{jm}, \nonumber
\end{eqnarray}
is used to diagonalize  the Hamiltonian. The eigenvalue problem may be expressed as a $2(N+1) \times 2(N+1)$ matrix,
{\small
\begin{eqnarray}
\left(\begin{array}{ccccccc}
\omega_m+2D_m-\omega & -2D_m & -i C_{1m} &  -iC_{1m} & \cdots & -i C_{Nm} &  -iC_{Nm} \\
2D_m & -\omega_m-2D_m-\omega & -i C_{1m} & -i C_{1m} & \cdots & -i C_{Nm} &  -iC_{Nm} \\
iC_{1m} & -i C_{1m} & \omega_{TO_1}-\omega &  0 & \cdots & 0 &  0 \\
-i C_{1m} & iC_{1m} & 0 &  -\omega_{TO_1}-\omega & \cdots & 0 &  0 \\
\vdots & \vdots & \vdots & \vdots & \vdots & \vdots & \vdots \\
iC_{Nm} & -i C_{Nm} & 0 &  0 & \cdots & \omega_{TO_N}-\omega &  0 \\
-i C_{Nm} & iC_{Nm} & 0 &  0 & \cdots & 0 &  -\omega_{TO_N}-\omega
\end{array} \right) \cdot \mathbf{F}=0. \nonumber
\end{eqnarray}
}
where $\mathbf{F}=(w_{mi}, y_{mi}, x_{1mi}, z_{1mi},\dots,x_{Nmi}, z_{Nmi})^t$ is a column vector, and
\begin{eqnarray}
D_m &=&\sum^N_{i=1} D_{im}=\frac{1}{4\omega_m} \sum^N_{i=1} \omega^2_{p_i}, \nonumber \\
C_{im} &=& \frac{\omega_{p_i}}{2} \sqrt{\frac{\omega_{TO_i}}{\omega_m}}, \nonumber
\end{eqnarray}
are the coupling constants. The secular equation may be also written as in the classical approach, $\omega^2 \epsilon(\omega)=\epsilon_\infty \omega^2_k $, where the dielectric constant $\epsilon(\omega)$ is a sum of several Lorentzian functions (Ec. \ref{eq:epswmulti}).

Employing the normalization condition, 
\begin{eqnarray}
w^*_{mi} w_{mi'}+\sum_j x^*_{jmi} x_{jmi'}-y^*_{mi} y_{mi'}-\sum_j z^*_{jmi} z_{jmi'}=\delta_{i,i'} \nonumber
\end{eqnarray}
we find the final expression for the normalized Hopfield coefficients for the eigenvalue $\omega_i$ of the $i^{th}$ polaritonic branch,
\begin{eqnarray}
w_{mi}  &=& \frac{1}{D}\left(1-\frac{\omega_i^2}{\omega^2_{TO_1}}\right)  \left(\frac{\omega_m}{\omega_{TO_1}}+\frac{\omega_i}{\omega_{TO_1}}\right), \nonumber \\
x_{jmi} &=& -\frac{i}{D}   \frac{1-\frac{\omega_i}{\omega_{TO_1}}}{1-\frac{\omega_i}{\omega_{TO_j}}} 
\frac{\omega_{p_i}}{\omega_{TO_j}} \sqrt{\frac{\omega_{TO_i} \omega_m}{\omega^2_{TO_1}}} \left(1+\frac{\omega_i}{\omega_{TO_1}}\right), \nonumber 
\end{eqnarray}
\begin{eqnarray}
 y_{mi} &=& - \frac{1}{D} \left(1-\frac{\omega_i^2}{\omega^2_{TO_1}}\right)  \left(\frac{\omega_m}{\omega_{TO_1}}-\frac{\omega_i}{\omega_{TO_1}}\right), \nonumber \\
z_{jmi} &=& -\frac{i}{D} \frac{1+\frac{\omega_i}{\omega_{TO_1}}}{1+\frac{\omega_i}{\omega_{TO_j}}}  \frac{\omega_{p_i}}{\omega_{TO_j}} \sqrt{\frac{\omega_{TO_i} \omega_m}{\omega^2_{TO_1}}} \left(1-\frac{\omega_i}{\omega_{TO_1}}\right), \nonumber 
\end{eqnarray}
which share the denominator
\begin{eqnarray}
D= 2 \left[\frac{\omega_m \omega_i}{\omega_{TO_1}^2}\left(\frac{\omega^2_{p_1}}{\omega^2_{TO_1}} +\left(1-\frac{\omega_i^2}{\omega_{TO_1}^2}\right)^2+\sum_{j>1} \frac{\omega^2_{p_j}}{\omega^2_{TO_j}} \frac{\left(1-\frac{\omega_i^2}{\omega_{TO_1}^2}\right)^2}{\left(1-\frac{\omega_i^2}{\omega_{TO_j}^2}\right)^2} \right)\right]^{1/2} \nonumber.
\end{eqnarray}
We have highlighted the relevance of the first vibrational mode for the experiemnts. All frequencies are normalized by $\omega_{TO_1}$. In Fig.  \ref{fig:ZP2}(a) the measured polaritonic frequencies are fitted with our model assuming that two vibrational modes interact with the cavity photon. The  corresponding Hopfield coefficients $|w-y|^2$ and $|x-z|^2$, depicted in Fig. \ref{fig:ZP2}(b), evaluate the relative contribution of photon and phonon modes to the polaritonic states. 
  
\section{Bulk phonon polaritons}
\label{sec:bphp}
Transverse waves in a homogeneous material also satisfy the propagation condition $\epsilon(\omega) \omega^2=\omega^2_k$, where $\omega_k=c k$, $c$ is the speed of light and $k$ the wavenumber of the plain wave. Following the approach outlined in Sec. \ref{sec:phonpol}, it is straightforward to show that bulk polaritonic modes are solution of the matrix equation (\ref{eq:multimatrix}). Fig. \ref{fig:bulkdr} shows the polaritonics bands in bulk SiO$_2$ for three vibrational modes.

\section{Fitting experimental data to the cavity model}
\label{sec:fit}
For a given coaxial array, filled with a phononic material, let us consider we have the spectral positions of the transmission resonances, $\omega_i$, for the $N + 1$ polaritonic branches corresponding to the  $N$ vibrational modes and the fundamental waveguide mode. The fitting to the polaritonic dispersion relation, $\omega^2 \epsilon(\omega)=\omega^2_k$, consists in the following two steps:
\begin{enumerate}
\item Extract the dispersion relation of the unfilled hole from the spectral positions: $\omega_k=\omega_{i_0} \sqrt{\epsilon(\omega_{i_0})}$. Notice that a single branch $i_0$ is needed. 
\item Solve the polynomial equation $\omega^2 \epsilon(\omega)=\omega_k$ for the remaining values of $i$ in order to obtain the fitting frequencies.
\end{enumerate}
Fig. \ref{fig:fit}(a) shows the results of the fitting for a coaxial gap of 21 nm filled with SiO$_2$. Fitting with one and three vibrational modes are compared with experimental results. 

\section{Tables}
\begin{table}[H]
\begin{center}
\begin{tabular}{c|c|c|c|c|c} \hline \hline
\multicolumn{2}{c|}{Gap (nm)}& 2 & 7 & 14 & 21 \\ \hline
\multirow{2}{*}{Bare ENZ mode} & Diameter (nm) & 520 & 520 & 550 & 550 \\ 
& Linewidth (cm$^{-1}$) (Q factor) & 453 (3.8) & 619 (3.6) & 844 (2.9) & 990 (2.7)  \\ \hline
\multirow{2}{*}{At zero-detuned} & Diameter (nm) & 580 & 790 & 940 & 940 \\ 
& Linewidth (cm$^{-1}$) (Q factor) & 245 (5.4) & 312 (4.5) & 438 (3.3) & 483 (3.0)  \\ \hline
    \multicolumn{2}{c|}{Mode splitting (cm$^{-1}$)}& 534 & 546 & 549 & 537 \\ \hline \hline
\end{tabular}
\end{center}
\caption{Measured linewidths of resonances shown in Figs S2 and S3}
\label{table:linewidths}
\end{table} 

\begin{table}[H]
\begin{center}
\begin{tabular}{c|c|c|c|c} \hline \hline
SiO$_2$ gap (nm) & 2 & 7 & 14 & 21 \\ \hline
$\eta$ & 0.508 & 0.519	& 0.522 & 0.510 \\
$D_m/\omega_{TO}$ (\%) & 6.45 & 6.73& 6.81 & 6.50 \\
$\Delta \omega^{cor}_{ZP}/\omega_{TO}$ (\%) & -3.33 & -3.48 & -3.52 & -3.35 \\ \hline
$\Delta \omega_{ZP}/\omega_{TO}$ (\%) & 3.12 & 3.25 & 3.29 & 3.15 \\
\hline \hline
\end{tabular}
\end{center}
\caption{Differential zero-point frequency, $\Delta \omega_{ZP}=D_m+\Delta \omega^{cor}_{ZP}$, at resonance, $\omega_k=\omega_{TO}$, for the measured coupling strength $\eta$. The negative correlation contribution, $\Delta \omega^{cor}_{ZP}$, and the positive coupling coefficient, $D_m$, are also reported. These quantities are normalized by the resonance frequency $\omega_{TO}$.}
\label{table:ZOE}
\end{table} 

\section{Figures}

\begin{figure}[H]
\centering
\includegraphics[scale=0.6]{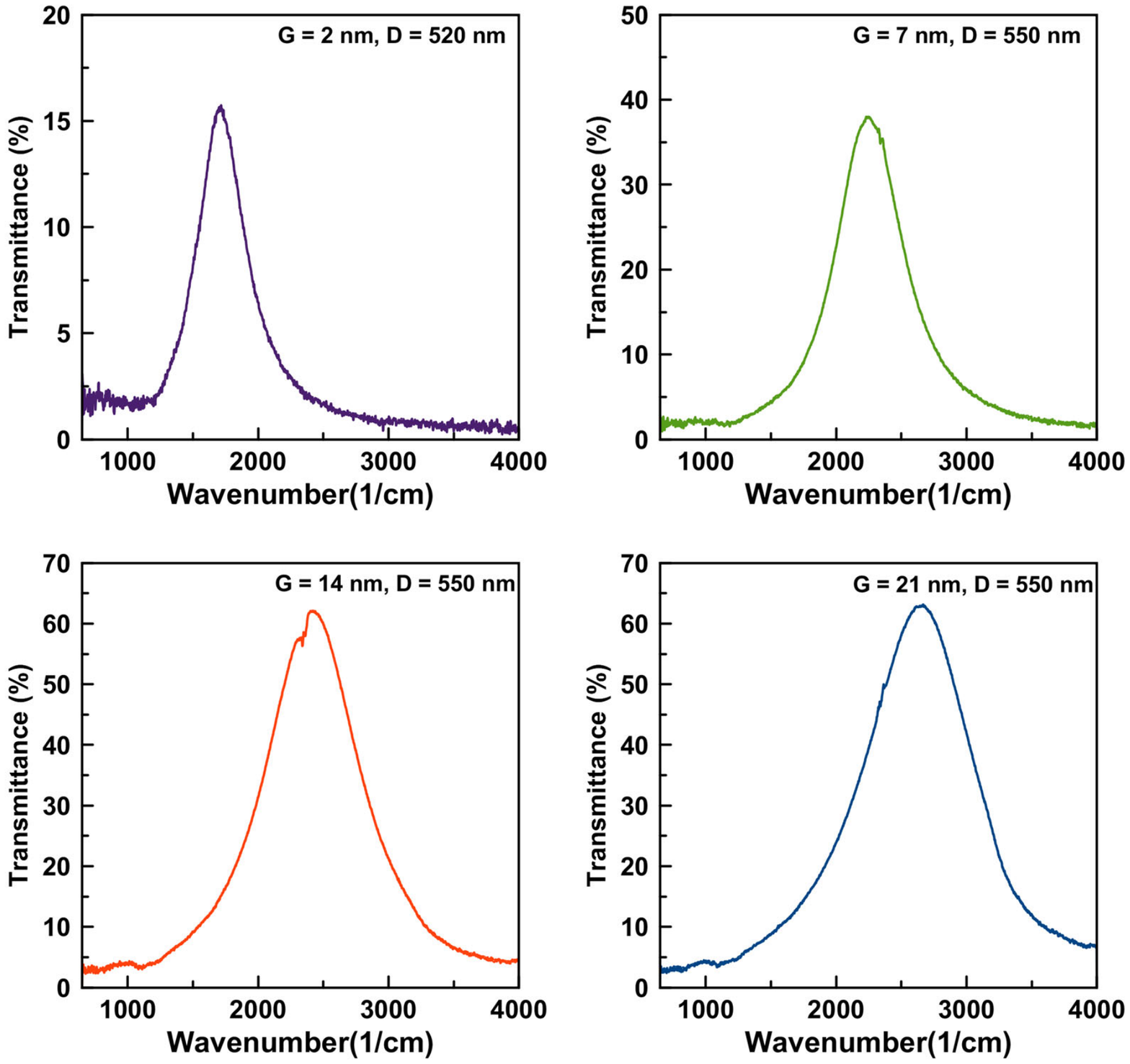}
\caption{Resonance spectra used for measuring the linewidth of bare ENZ resonance.}
\label{fig:USC_SI_Fig_S5}
\end{figure}

\begin{figure}[H]
\centering
\includegraphics[scale=0.6]{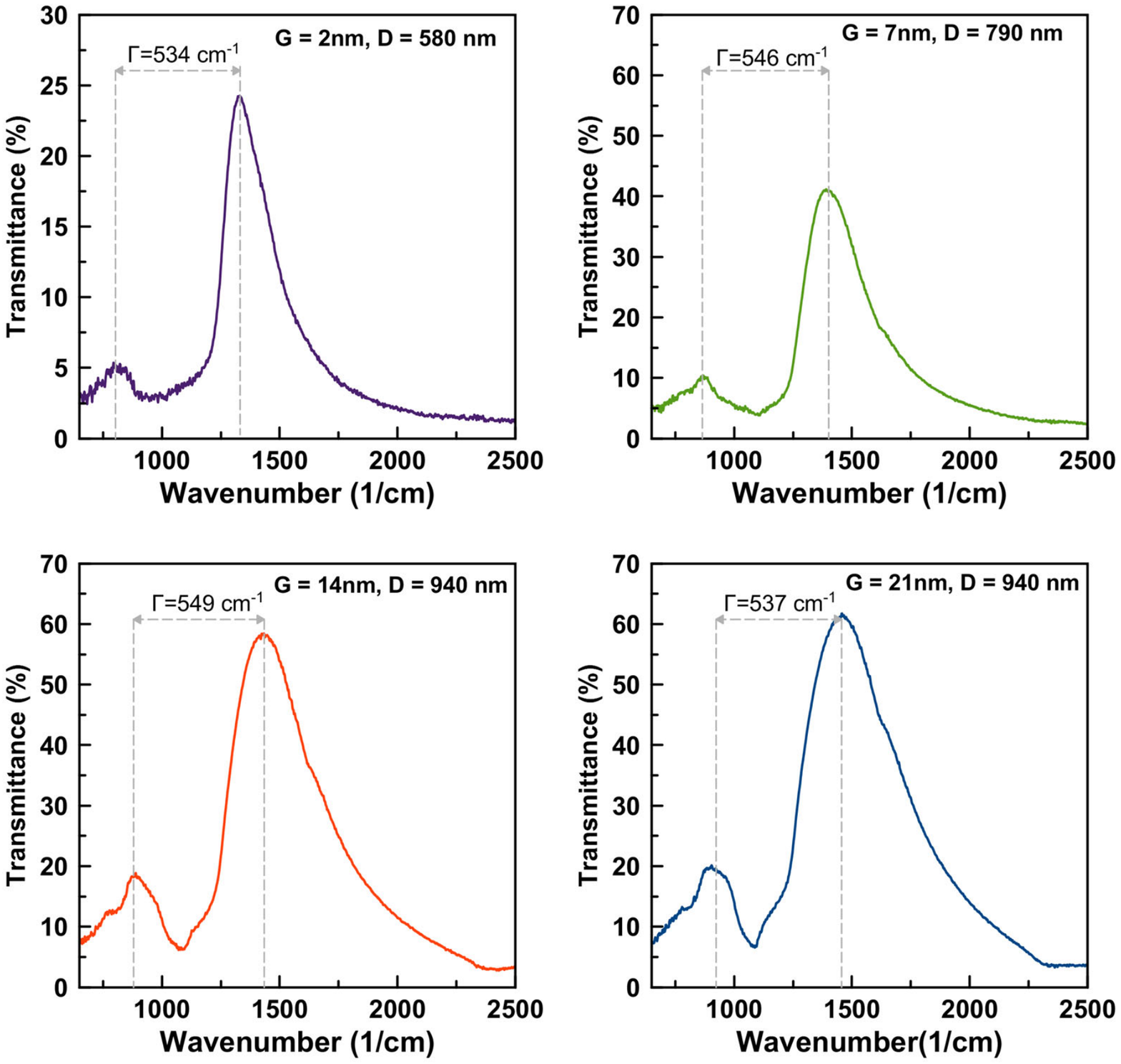}
\caption{Resonance spectra used for measuring both the linewidth of coupled ENZ resonance and the resulting mode splitting.}
\label{fig:USC_SI_Fig_S6}
\end{figure}

\begin{figure}[H]
\centering
\includegraphics[scale=0.35]{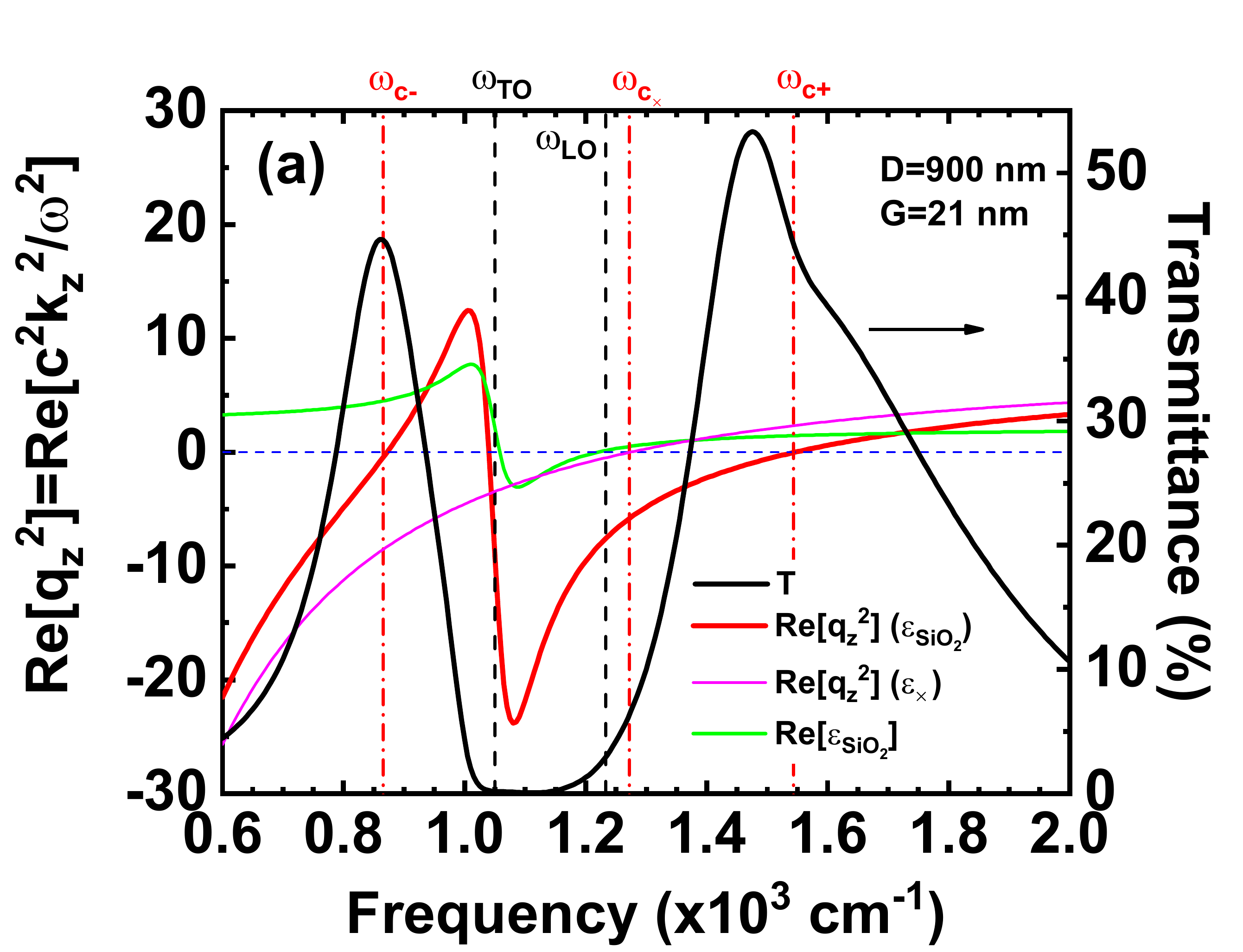}
\includegraphics[scale=0.35]{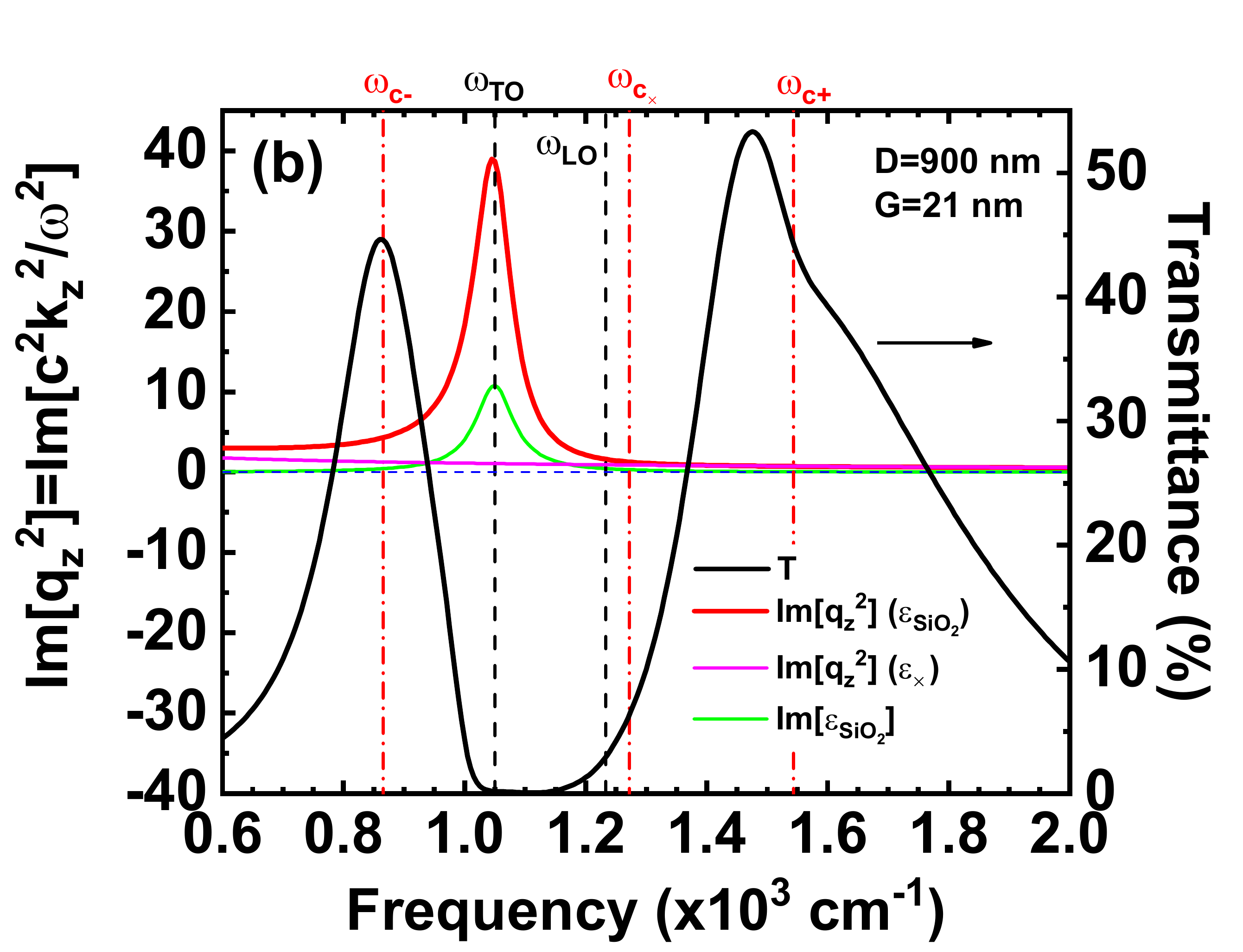}
\caption{(a) Real and (b) imaginary parts of the squared effective index, $q^2_z$ (red line), of the fundamental mode TE$_{11}$ as a function of a frequency for a SiO$_2$ filled coaxial gap with diameter $D=900$ nm and gap size $G=21$ nm. This quantity is compared with $q^2_z$ for a $\epsilon_\infty$ filled gap (magenta line) and the dielectric constant of SiO$_2$ (green line). The cutoff frequencies (red dash-dotted vertical lines) of the SiO$_2$ filled gap are computed with Eq. (\ref{eq:cutoff}). The transmittance for a hexagonal hole array with 1190 nm period, and 80 nm Au thickness (black line) shows peaks when hole are at cutoff, i.e. for Re$[q^2_z]=0$. The coupled-mode method was employed for computing the transmission spectrum \cite{FJRMP10}.}
\label{fig:cutoff}
\end{figure}

\begin{figure}[H]
\centering
\hspace{-6.5cm}
\includegraphics[scale=0.6]{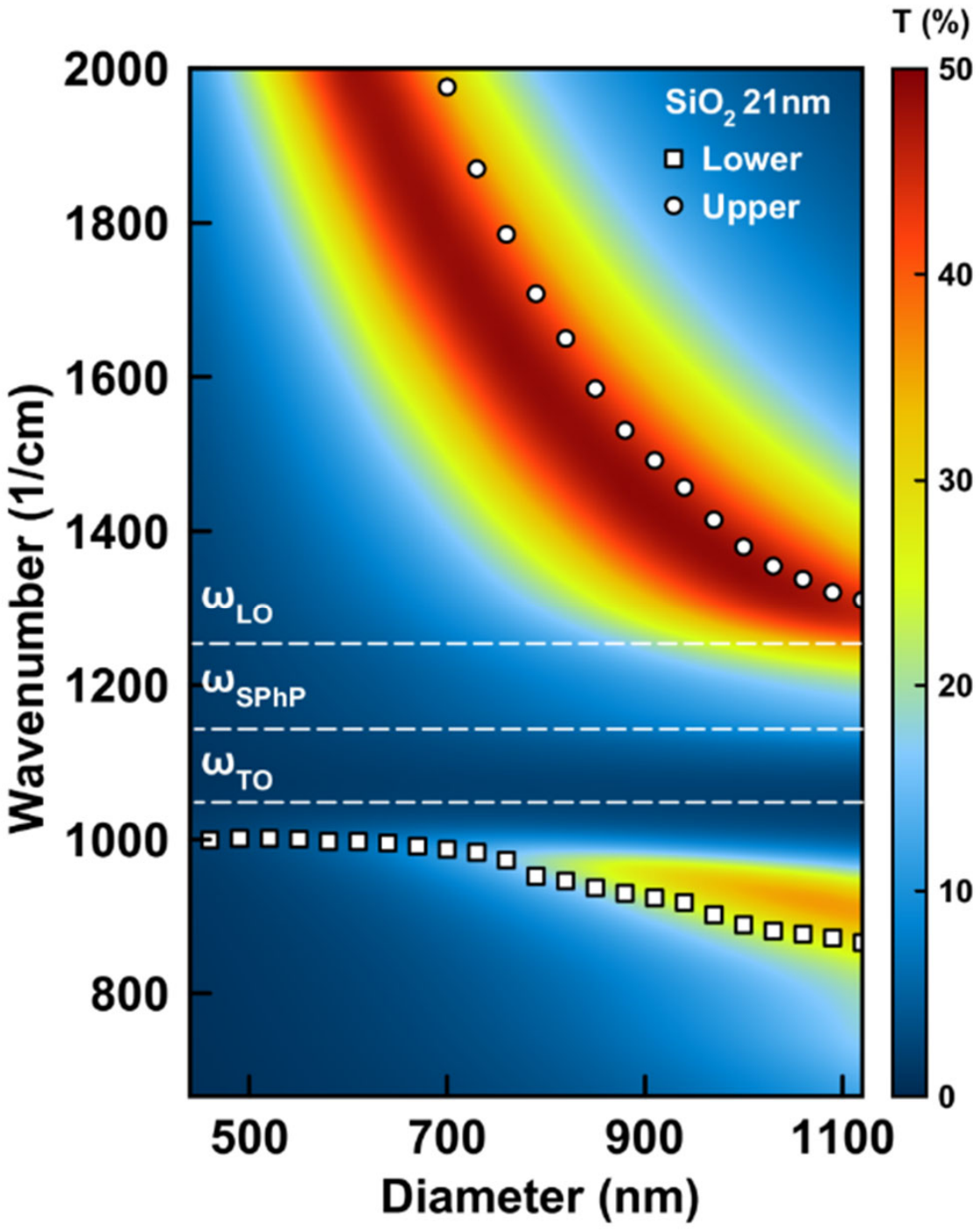}
\caption{Color map shows simulated transmission spectra of coaxial nanoapertures (21 nm SiO$_2$ gap) in a gold film as a function of the coax diameter. White circular (square) dots indicate the upper (lower) branch spectral peak positions measured from FTIR spectra in Fig. 2b.}
\label{fig:Tcountourplot}
\end{figure}

\begin{figure}[H]
\centering
\includegraphics[scale=0.4]{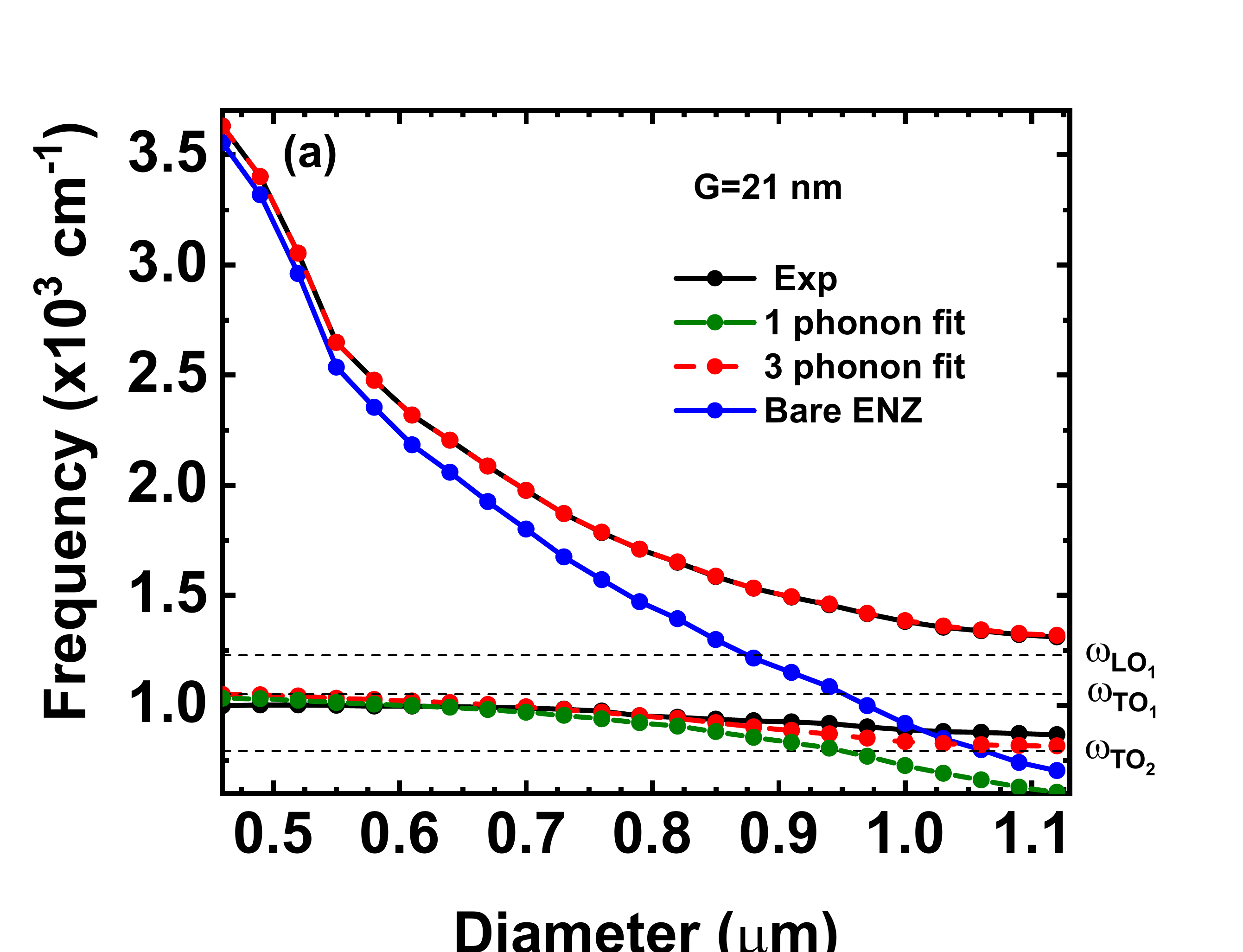}
\includegraphics[scale=0.4]{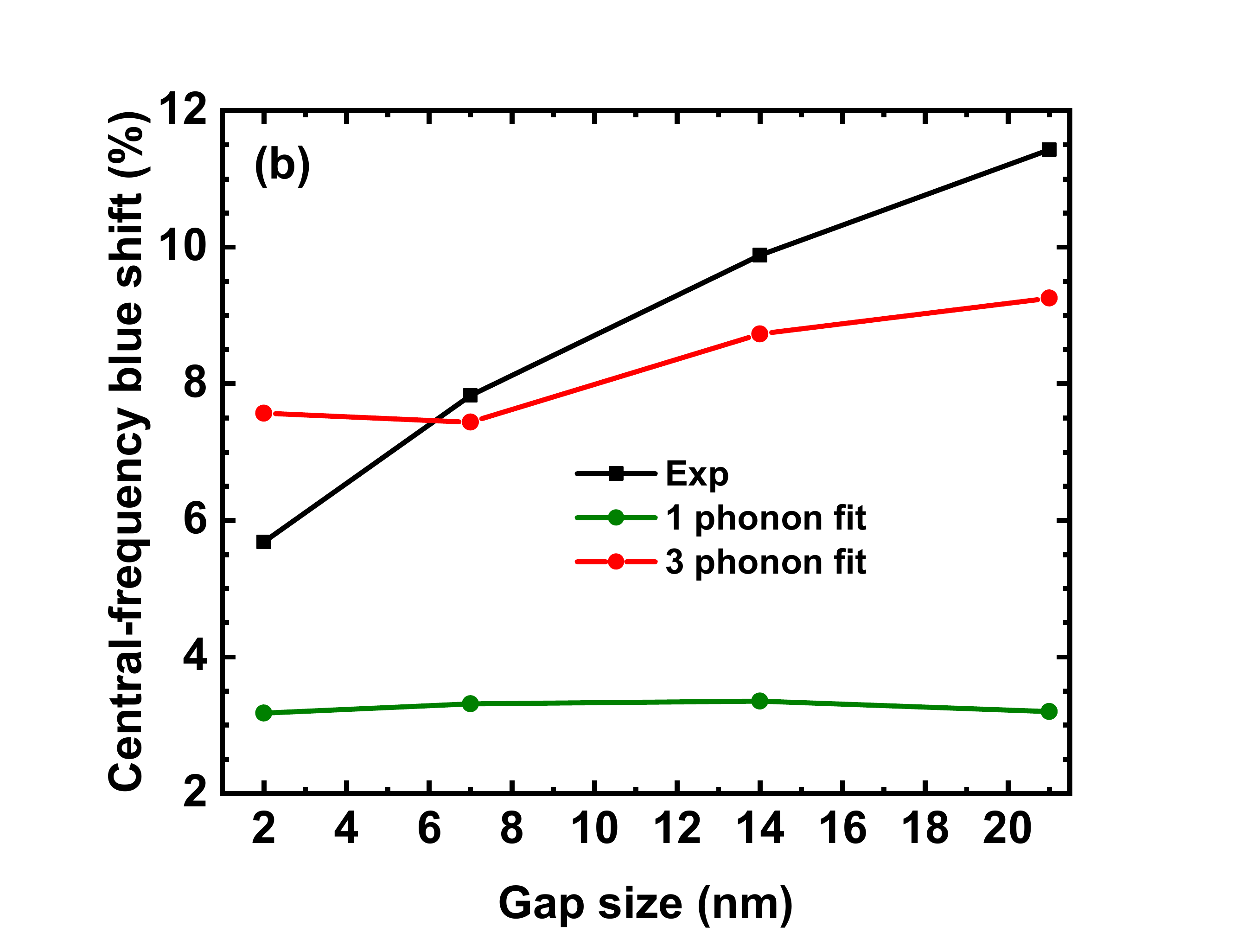}
\caption{(a) Experimental peak positions (black) for a coaxial gap of 21 nm filled with SiO$_2$ are compared with the fitted values for 1 (green) and 3 (red) vibrational modes as well as bare ENZ modes (blue). (b) Blue shifting, $\Delta \omega_{cen}=(\omega_{cen}-\omega_{TO_1})/\omega_{TO_1}$ $[\times 100\%]$, of the central frequency, $\omega_{cen}=(\omega_++\omega_-)/2$ at resonance ($\omega_{ENZ}=\omega_{TO_1}$).}
\label{fig:fit}
\end{figure}

\begin{figure}[H]
\centering
\includegraphics[scale=1.0]{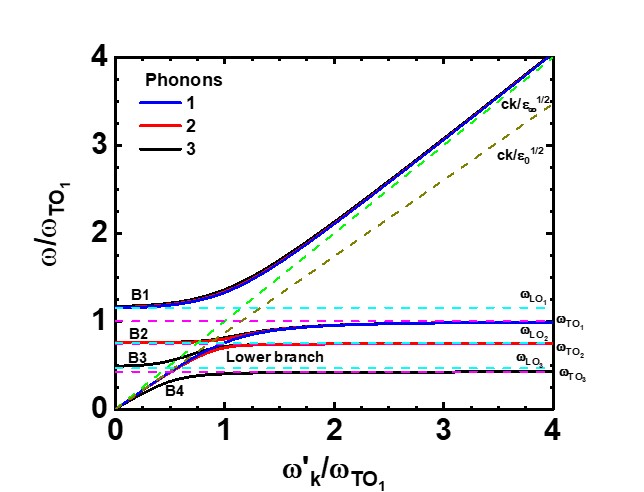}
\caption{Dispersion relation of phonon polaritons for the vibrational modes of bulk SiO$_2$.}
\label{fig:bulkdr}
\end{figure}

\begin{figure}[H]
\centering
\includegraphics[scale=1.0]{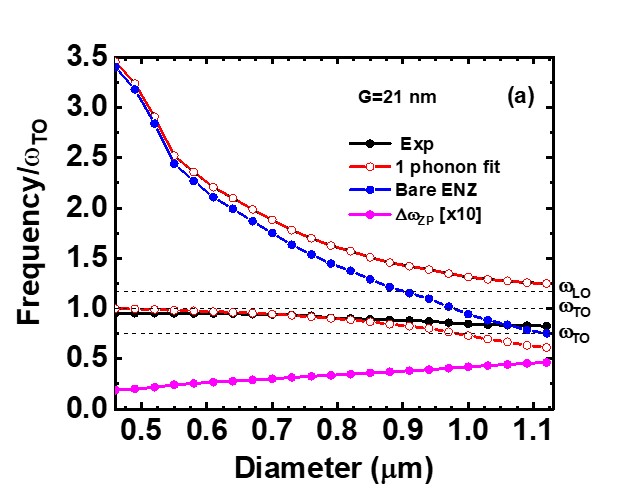}
\includegraphics[scale=1.0]{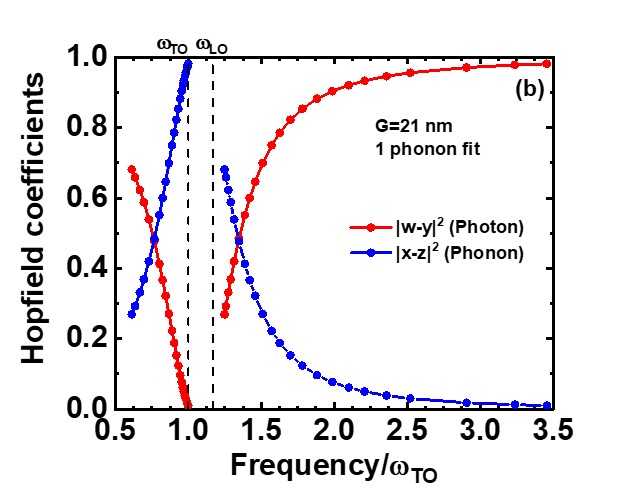}
\caption{({\bf Color online}). (a) Measured phonon polariton branches  are compared with the calculated zero-point differential frequency, $\Delta \omega_{ZP}[x10]$. Experimental points are fitted with the  eigenfrequencies calculated assuming that cavity photons interact with a single vibrational mode of SiO$_2$. The bare ENZ frequency is the single fitting parameter. (b) Hopfield coefficients corresponding to the measured frequencies.}
\label{fig:ZP1}
\end{figure}

\begin{figure}[H]
\centering
\includegraphics[scale=1.0]{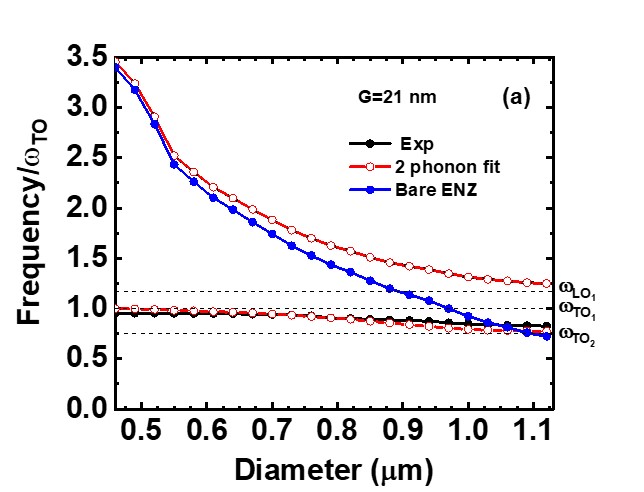}
\includegraphics[scale=1.0]{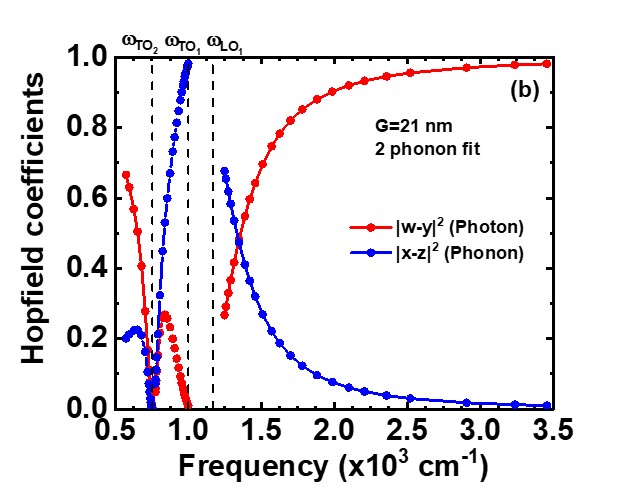}
\caption{({\bf Color online}). (a) Measured phonon polariton branches   fitted with the  eigenfrequencies calculated assuming that cavity photons interact with two vibrational modes of SiO$_2$. The bare ENZ frequency is the single fitting parameter. (b) Hopfield coefficients corresponding to the measured frequencies.}
\label{fig:ZP2}
\end{figure}

\bibliography{references}
\bibliographystyle{Science}

\end{document}